\documentclass[12pt]{article}
\usepackage{amsfonts}
\usepackage{amssymb}
\usepackage{graphicx}
\begin{document}
\renewcommand{\thefootnote}{\fnsymbol{footnote}}
\begin{titlepage}
\begin{flushright}
\end{flushright}

\vspace{6mm}

\begin{center}
{\Large\bf Classical Mechanics on Noncommutative Space with Lie-algebraic Structure}

\vspace{13mm}

{{\large Yan-Gang Miao${}^{1,2,3,}$\footnote{{\em E-mail: miaoyg@nankai.edu.cn}},
Xu-Dong Wang${}^{1,}$\footnote{{\em E-mail: c\_d\_wang@mail.nankai.edu.cn}},
and Shao-Jie Yu${}^{1,}$\footnote{{\em E-mail: leptonyu@gmail.com}}}\\

\vspace{6mm}
${}^{1}${\em School of Physics, Nankai University, Tianjin 300071, \\
People's Republic of China}

\vspace{3mm}
${}^{2}${\em Department of Physics, University of Kaiserslautern, P.O. Box 3049,\\
D-67653 Kaiserslautern, Germany}

\vspace{3mm}
${}^{3}${\em Bethe Center for Theoretical Physics and Institute of Physics, University of Bonn, \\
Nussallee 12, D-53115 Bonn, Germany}}

\end{center}

\vspace{8mm}
\centerline{{\bf{Abstract}}}
\vspace{6mm}

We investigate the kinetics of a nonrelativistic particle interacting with a constant external force on a
Lie-algebraic noncommutative space. The structure constants of a Lie algebra, also called noncommutative parameters,
are constrained in general due to some algebraic properties, such as the antisymmetry and Jacobi identity.
Through solving the constraint equations
the structure constants satisfy, we obtain two new sorts of algebraic structures,
each of which corresponds to one type of noncommutative spaces.
Based on such types of noncommutative spaces as the starting point,
we analyze the classical motion of the particle interacting with a constant external force by means of the Hamiltonian formalism on a Poisson manifold.
Our results {\em not only} include that of
a recent work as our special cases, {\em but also} provide new trajectories of motion
governed mainly by marvelous extra forces.
The extra forces with the unimaginable $t\dot{x}$-, $\dot{(xx)}$-, and $\ddot{(xx)}$-dependence
besides with the usual $t$-, $x$-, and $\dot{x}$-dependence, originating from a variety of noncommutativity
between different spatial coordinates and between spatial coordinates and momenta as well,
deform greatly the particle's ordinary trajectories we are quite familiar with on the Euclidean (commutative) space.

\vskip 12pt
PACS Number(s): 11.10.Nx; 45.50.Dd; 45.20.da


\end{titlepage}

\newpage
\renewcommand{\thefootnote}{\arabic{footnote}}
\setcounter{footnote}{0}
\setcounter{page}{2}

\section{Introduction}

Physics founded on noncommutative spacetimes has developed rapidly since the end of last century
when the idea of
spacetime noncommutativity~\cite{s1} was revived in the field of string theory~\cite{s2}.
There has been a large
amount of literature on noncommutative quantum mechanics (NCQM)~\cite{s3} and noncommutative field
theory (NCFT)~\cite{s4} as well. However, we notice that the research
on noncommutative classical mechanics\footnote{Here the classical mechanics includes
both the nonrelativistic and relativistic mechanics.} (NCCM)~\cite{s5,s6,s7},
if comparing with that of NCQM and NCFT, is so little. Probably the NCCM is not so attractive as the NCQM and NCFT;
nevertheless, the Doubly Special Relativity~\cite{s7} has been intriguing. Moreover,
the present paper tries to give from the nonrelativistic aspect a glance at a variety of interesting properties that the NCCM possesses.

The mathematical background for the physics on noncommutative spacetimes is the noncommutative geometry~\cite{s8}.
As was demonstrated, {\em e.g.} in Ref.~\cite{s5}, the spacetime noncommutativity can be distinguished
into three kinds in accordance with the Hopf-algebraic classification~\cite{s9}, that is, there exist
the canonical, Lie-algebraic and quadratic noncommutativity, respectively.
In addition, the three types of noncommutative spacetimes have been studied in the framework of quantum groups at
both the nonrelativistic and relativistic levels, and the relative Hopf algebras
for some specific noncommutative spacetimes\footnote{For example, the $\kappa$-deformed
Minkowski spacetime~\cite{s14} as a specific case of the Lie-algebraic noncommutativity
has recently been paid much attention because it is a natural candidate
for the spacetime based on which the Doubly Special Relativity~\cite{s7} has been established.}
have been given~\cite{s10,s11,s12,s13,s14,s15}. In brief,
at the former level the Galilei Hopf algebras have been provided for the
canonical~\cite{s10,s12}, Lie-algebraic~\cite{s10,s11,s12} and quadratic~\cite{s12} noncommutativity, respectively;
and at the latter level the Poincar$\acute{\rm e}$ Hopf algebras have been proposed for the
canonical~\cite{s13,s10}, Lie-algebraic~\cite{s14,s10,s11} and quadratic~\cite{s15} noncommutativity, respectively.
Incidentally, the Hopf algebras for general noncommutative spacetimes still remain
unknown.\footnote{For instance, it may be a possible progress to construct the Hopf algebras for
Type I and Type II spaces (eqs.~(8) and (9)).}

In a recent work~\cite{s5} a nonrelativistic particle interacting with a constant external force on the (Lie-algebraic and quadratic) noncommutative
phase spaces with commutative momenta was dealt with in detail and some intriguing trajectories were then revealed.
This classical system is independent of any star-products and can thus be analyzed
by means of the Hamiltonian formalism on a Poisson manifold. The reason lies both on the constant external force and on the commutative momenta
(see the discussions in the next section for the details).
If the external force were not constant and the momenta were not commutative,
one would have to envisage the corresponding star-product rules which are complicated and unknown at present.
Therefore, Ref.~\cite{s5} provides an alternative way to study a simplified classical system on a kind of specific (Lie-algebraic and quadratic)
noncommutative spaces of which the rules of star-products are unknown. In other words, one is able to analyze some simple classical motion,
such as that associated with a constant external force, on  so complicated noncommutative spaces that their corresponding star-product rules are
even unknown.
However, we would like to point out that the (Lie-algebraic) noncommutative spaces considered in Ref.~\cite{s5}
as the starting point for investigating the particle's noncommutative kinetics
are special cases of our findings.

In this paper we generalize Ref.~\cite{s5}
to more complicated Lie-algebraic noncommutative spaces.
Our generalization focuses only on the Poisson brackets between different spatial coordinates
and between spatial coordinates and momenta, but still keeps the Poisson brackets between different momenta vanishing.
See eqs.~(8) and (9). That is, we consider
our generalization within the limitation that no star-products are needed, which can be seen clearly in
the Newton equations, eqs.~(23)-(25) and eqs.~(38)-(40), where no products of spatial coordinates
associated with non-vanishing Poisson brackets exist.
To this end, we propose a practical and effective way to look for more complicated Lie-algebraic noncommutative spaces than that appeared
in Ref.~\cite{s5},
which is in fact based on Lie-algebraic properties,
such as the antisymmetry and Jacobi identity. In other words, we
solve the constraint equations that the structure constants of a Lie algebra,
here known as noncommutative parameters, hold,
and obtain two new types of noncommutative spaces with the Lie-algebraic structure which cover that of Ref.~\cite{s5} as subclasses.
In terms of the Hamiltonian analysis
on a Poisson manifold~\cite{s16}, we derive the Hamilton equation and Newton equation, and
observe that the noncommutativity between different spatial coordinates and between spatial coordinates and momenta
provides various extra forces that
make the particle's ordinary trajectories on the Euclidean (commutative) space have a quite large deformation.
We point out, as a byproduct, that the Lie-algebraic noncommutative spaces are anisotropic because of the
direction-dependent deformation.

The arrangement of this paper is as follows. In the next section,
we determine new types of noncommutative spaces with the Lie-algebraic structure.
In section 3, we study the kinetics of a nonrelativistic particle interacting with a
constant external force on the new noncommutative spaces given by section 2.
Finally section 4 is devoted to the conclusion and perspective.

\section{Lie-algebraic noncommutativity}

We start with general Poisson brackets of spatial coordinates of a noncommutative space with the Lie-algebraic
structure,\footnote{It is no doubt that the repeated indices stand for summation.}
\begin{equation}
\{x_a, x_b\}={\theta}^0_{ab}t+{\theta}^c_{ab}x_c,
\end{equation}
where lowercase Latin indices $a, b, \cdots, e=1, 2, 3$, time $t$ is dealt with as a parameter which
has vanishing Poisson brackets with
spatial coordinates, and the structure constants, also called noncommutative parameters,
${\theta}^0_{ab}$ and ${\theta}^c_{ab}$ are antisymmetric to lower indices
and, moreover, have to satisfy some quadratic relations, {\em i.e.} the constraint equations due to the Jacobi
identity as follows:
\begin{eqnarray}
{\theta}^d_{ab}{\theta}^0_{dc}+{\theta}^d_{ca}{\theta}^0_{db}+{\theta}^d_{bc}{\theta}^0_{da}&=&0,\nonumber \\
{\theta}^e_{ab}{\theta}^d_{ec}+{\theta}^e_{ca}{\theta}^d_{eb}+{\theta}^e_{bc}{\theta}^d_{ea}&=&0.
\end{eqnarray}
Furthermore, we implement the momentum space by imposing the {\em ad hoc} Poisson brackets upon eq.~(1),
\begin{eqnarray}
\{x_a, p_b\}&=&{\delta}_{ab} + {\bar{\theta}}^c_{ab}x_c + {\tilde{\theta}}^c_{ab}p_c, \\
\{p_a, p_b\}&=&0,
\end{eqnarray}
where ${\bar{\theta}}^c_{ab}$ and ${\tilde{\theta}}^c_{ab}$ are newly introduced noncommutative parameters that have
vanishing diagonal elements. As a result,
while maintaining a quite some generality for the noncommutative phase space we still require
eq.~(3) to recover the usual canonical relations, {\em i.e.} the Poisson bracket version of
the standard Heisenberg commutation relations if $a=b$. As mentioned in the above section,
the vanishing Poisson brackets of momenta are chosen in order to avoid highly nontrivial differential
calculus related to star-products~\cite{s17}, and
such a simplicity in the recent studies of noncommutativity~\cite{s3,s4,s6,s7} is acceptable due to a variety of
nonvanishing Poisson brackets that have been introduced between different spatial coordinates
and between spatial coordinates and momenta
as well. In the phase space spanned by $(x_a,p_a)$, the Jacobi identity gives rise to, besides eq.~(2), additional
(linear and quadratic) constraint equations the noncommutative parameters should hold,
\begin{eqnarray}
& &{\bar{\theta}}^a_{cb}-{\bar{\theta}}^b_{ca}=0,\nonumber \\
& &{\theta}^c_{ab}+{\tilde{\theta}}^b_{ac}-{\tilde{\theta}}^a_{bc}=0,\nonumber \\
& &{\bar{\theta}}^d_{bc}{\theta}^0_{da}-{\bar{\theta}}^d_{ac}{\theta}^0_{db}=0,\nonumber \\
& &{\bar{\theta}}^e_{ca}{\bar{\theta}}^d_{eb}-{\bar{\theta}}^e_{cb}{\bar{\theta}}^d_{ea}=0,\nonumber \\
& &{\bar{\theta}}^e_{ca}{\tilde{\theta}}^d_{eb}-{\bar{\theta}}^e_{cb}{\tilde{\theta}}^d_{ea}=0,\nonumber \\
& &{\theta}^e_{ab}{\tilde{\theta}}^d_{ec}-{\tilde{\theta}}^e_{bc}{\tilde{\theta}}^d_{ae}
+{\tilde{\theta}}^e_{ac}{\tilde{\theta}}^d_{be}=0,\nonumber \\
& &{\theta}^e_{ab}{\bar{\theta}}^d_{ec}+{\bar{\theta}}^e_{bc}{\theta}^d_{ea}-{\bar{\theta}}^e_{ac}{\theta}^d_{eb}
-{\tilde{\theta}}^e_{bc}{\bar{\theta}}^d_{ae}+{\tilde{\theta}}^e_{ac}{\bar{\theta}}^d_{be}=0,
\end{eqnarray}
where the antisymmetry of lower indices of ${\theta}^0$, ${\theta}$,
${\tilde{\theta}}$, and ${\bar{\theta}}$ should be considered. The
dimensions of the parameters are fixed: $[{\theta}^0]=[M]^{-1}$,
$[{\theta}]=[{\tilde{\theta}}]=[M]^{-1}[L]^{-1}[T]$, and
$[{\bar{\theta}}]=[L]^{-1}$, where $M$, $L$, and $T$ stand for mass,
length, and time, respectively. As a summary, we mention that the noncommutative space is governed by the Lie algebra which contains seven generators,
$t, x_1, x_2, x_3, p_1, p_2, p_3$, where $t$ is central
and $p_1,p_2,p_3$ form an abelian subalgebra. This Lie algebra leads to a Poisson manifold on $\mathcal{R}^6$ with coordinates
$x_1, x_2, x_3, p_1, p_2, p_3$
and its structure constants satisfy eqs.~(2) and (5).

In Ref.~\cite{s5}, however, two Lie-algebraic noncommutative spaces are simply supposed.
We shall show that they are specific kinds of solutions of eqs.~(2) and (5) and are covered by our solutions (see eqs.~(6) and (7) or eqs.~(8) and (9)).
The two sorts of algebraic structures of noncommutative spaces are summarized as follows:
\begin{itemize}
\item One of the three Poisson brackets of different spatial coordinates is deformed to be proportional
to time, but the other brackets of different spatial coordinates, of spatial coordinates and momenta,
and of different momenta maintain undeformed;
\item Two of the Poisson brackets of different spatial
coordinates are deformed to be proportional to a spatial coordinate and
two of the Poisson brackets of spatial coordinates and momenta are deformed to be proportional to a momentum,
but the other brackets maintain undeformed.
\end{itemize}
Based on the above two noncommutative spaces, the kinetics
of a nonrelativistic particle interacting with a constant external force is discussed and
some interesting trajectories of motion are revealed.

Nevertheless, we point out
that the constraint equations, eqs.~(2) and (5), contain far more
plentiful structures of noncommutative spaces than that given by
Ref.~\cite{s5}. It is obviously a difficult task to solve the constraint equations, {\em i.e.} eqs.~(2) and (5) which contain so many related equations
and have to comply with the antisymmetry of lower indices. This presents that the classical mechanics does not always mean simple, in particular
when it is associated with noncommutativity.
In the present paper,
after tedious calculations together with some calculating techniques, we at last obtain two kinds of solutions from eqs.~(2) and (5),
each of which corresponds to
a new noncommutative space with a typical Lie-algebraic structure on which the classical mechanics would be established.
Although they are not the most general solutions,
the two kinds of solutions we work out are at least the
generalizations of that of Ref.~\cite{s5} and in particular lead to various marvelous
classical trajectories unknown before, see the next section in detail.


The two kinds of solutions are listed
as follows:\footnote{For the sake of convenience of comparing with the results of Ref.~\cite{s5},
we use the same notation as that adopted there.}
\begin{itemize}
\item For the first kind of solutions, the non-vanishing elements of structure constants are
\begin{equation}
{\theta}^0_{kl}=-{\theta}^0_{k{\gamma}}=\frac{1}{{\kappa}}, \qquad
{\theta}^0_{l{\gamma}}=\frac{1}{{\kappa}}, \qquad
{\theta}^l_{k{\gamma}}=-{\theta}^k_{l{\gamma}}=\frac{1}{{\tilde{\kappa}}}, \qquad
{\tilde{\theta}}^l_{k{\gamma}}=-{\tilde{\theta}}^k_{l{\gamma}}=\frac{1}{{\tilde{\kappa}}},
\end{equation}
while the others are zero except for the antisymmetric counterparts of the above elements.
\item For the second kind of solutions, the non-vanishing elements of structure constants are
\begin{equation}
{\theta}^0_{l{\gamma}}=-{\theta}^0_{k{\gamma}}=\frac{1}{{\kappa}}, \qquad
{\theta}^l_{k{\gamma}}=-{\theta}^k_{l{\gamma}}=\frac{1}{{\tilde{\kappa}}}, \qquad
{\tilde{\theta}}^l_{k{\gamma}}=-{\tilde{\theta}}^k_{l{\gamma}}=\frac{1}{{\tilde{\kappa}}},\qquad
{\bar{\theta}}^l_{k{\gamma}}={\bar{\theta}}^k_{l{\gamma}}=\frac{1}{{\bar{\kappa}}},
\end{equation}
while the others are zero except for the antisymmetric counterparts of the above elements.
\end{itemize}
Note that $k, l$, and $\gamma$ (running over $1, 2, 3$) are different and fixed,
and that ${\kappa}$, ${\tilde{\kappa}}$, and ${\bar{\kappa}}$
are independent noncommutative parameters whose dimensions are $[{\kappa}]=[M]$,
$[{\tilde{\kappa}}]=[M][L][T]^{-1}$, and $[{\bar{\kappa}}]=[L]$, respectively.

Type I noncommutative space, which corresponds to the first kind of solutions (eq.~(6)) and
generalizes the first Lie-algebraic noncommutative space given by Ref.~\cite{s5}, has the form,
\begin{equation}
\begin{array}{llll}
\{x_k,x_{\gamma}\}=-\frac{1}{\kappa}t+\frac{1}{\tilde{\kappa}}x_l, &  \{x_l,x_{\gamma}\}
=\frac{1}{\kappa}t-\frac{1}{\tilde{\kappa}}x_k,  &
\{x_k,x_{l}\}=\frac{1}{\kappa}t; \\
\{p_k,x_{\gamma}\}=\frac{1}{\tilde{\kappa}}p_l, &
\{p_l,x_{\gamma}\}=-\frac{1}{\tilde{\kappa}}p_k, \\ \{x_i,p_j\}={\delta}_{ij},& \{x_{\gamma},p_{\gamma}\}=1;\\
\{p_a,p_b\}=0,
\end{array}
\end{equation}
where $i,j=k,l$, and $a,b=k, l, \gamma$.
When ${\tilde{\kappa}} \rightarrow \infty$ and ${\theta}^0_{kl}={\theta}^0_{k{\gamma}}=0$ in eq.~(6),
Type I reduces to the first Lie-algebraic formulation of Ref.~\cite{s5}.

Type II noncommutative space, which corresponds to the second kind of solutions (eq.~(7)) and
generalizes the second Lie-algebraic noncommutative space given by Ref.~\cite{s5},
takes the form,
\begin{equation}
\begin{array}{llll}
\{x_k,x_{\gamma}\}=-\frac{1}{\kappa}t+\frac{1}{\tilde{\kappa}}x_l, &  \{x_l,x_{\gamma}\}
=\frac{1}{\kappa}t-\frac{1}{\tilde{\kappa}}x_k,  &
\{x_k,x_{l}\}=0; \\
\{p_k,x_{\gamma}\}=\frac{1}{\bar{\kappa}}x_{l}+\frac{1}{\tilde{\kappa}}p_l, &
\{p_l,x_{\gamma}\}=\frac{1}{\bar{\kappa}}x_{k}-\frac{1}{\tilde{\kappa}}p_k, \\ \{x_i,p_j\}={\delta}_{ij},&
\{x_{\gamma},p_{\gamma}\}=1;\\
\{p_a,p_b\}=0.
\end{array}
\end{equation}
When ${{\kappa}} \rightarrow \infty$ and ${\bar{\kappa}} \rightarrow \infty$ simultaneously in eq.~(7) or eq.~(9),
Type II reduces to
the second Lie-algebraic formulation of Ref.~\cite{s5}. Moreover, if ${{\kappa}} \rightarrow \infty$ and
${\tilde{\kappa}} \rightarrow \infty$ individually,
Type II is simplified to its two subclasses of noncommutative spaces
which are also novel and untouched in literature. As to the two subclasses, though included in Type II,
we may consider them separately elsewhere because they are related to interesting trajectories of motion.

We emphasize that eqs.~(8) and (9) indeed describe new noncommutative spaces with the Lie-algebraic structure rather than
ordinary (commutative) spaces with noncanonical Poisson brackets.
This can be seen clearly when we notice that eqs.~(8) and (9) involve the two noncommutative spaces utilized in
Ref.~\cite{s5} as their subclasses while the latter have been proved~\cite{s9} to be recovered in a Hopf
algebraic framework with a nontrivial coalgebra. Or, from a different point of view,
as Darboux theorem holds only locally and not on all $\mathcal{R}^6$,
one cannot find canonical coordinates and their conjugate momenta that satisfy the Poisson bracket version of
the standard Heisenberg commutation relations through introducing a linear or nonlinear transformation
in such an original phase space $\mathcal{R}^6$ that is governed by a Hopf algebra with a nontrivial coalgebra.
As a consequence, the extra forces (see the following section) emerge indeed from the noncommutativity of the phase space $\mathcal{R}^6$
rather than from an unsuitable choice of noncanonical Poisson brackets.

At the end of this section we point out that the three (spatial or momental) directions in eqs.~(8) and (9)
are not equivalent to
each other, which results in the conclusion that the two Lie-algebraic noncommutative spaces are anisotropic.
This feature will be seen clearly through the deformation of the ordinary trajectories of motion with which we are quite familiar
on the Euclidean (commutative) space.

In the next section we turn to the study on the classical kinetics of a nonrelativistic particle interacting with
a constant external force
on the two types of Lie-algebraic noncommutative spaces depicted by eqs.~(8) and (9).

\section{Classical mechanics on Lie-algebraic noncommutative space}
In this section we analyze the trajectories of a nonrelativistic particle interacting with a constant external force,
$\vec{F}=(F_k, F_l, F_{\gamma})$, on the noncommutative
spaces Type I and Type II. As we shall see, the noncommutativity produces extra forces
that have dynamical effects and thus alter in the way of direction-dependence
the particle's ordinary trajectories, {\em i.e.} the trajectories on the Euclidean (commutative) space.
This phenomenon reflects that the two Lie-algebraic noncommutative spaces are anisotropic.

The problem we are now envisaging is to establish equations of motion for a particle on a Lie-algebraic
noncommutative space. That is, we have to generalize the canonical Hamilton equation related to the Euclidean
space to such a Hamilton equation that holds on a Lie-algebraic noncommutative space. In fact, this is only
a simple application of a Poisson manifold.

In order for this paper to be self-contained we briefly repeat the main context of the
Hamiltonian analysis on a Poisson manifold~\cite{s16}, but use the present notation for the sake of consistency in
the paper as a whole.

At first, we recall the definition of a Poisson manifold and the equation of motion on the Poisson manifold.

{\bf Definition}. {\em A {\bf Poisson bracket} (or a {\bf Poisson structure}) on a manifold $P$ is a bilinear
operation $\{ , \}$ on ${\mathcal F}(P)=C^{\infty}(P)$ such that:
\begin{enumerate}
\item $({\mathcal F}(P), \{ , \})$ is a Lie algebra; and
\item $\{ , \}$ is a derivation in each factor, that is,
\begin{equation}
\{FG, H\}=\{F, H\}G+F\{G, H\}
\end{equation}
for all $F$, $G$, and $H \in {\mathcal F}(P)$.
\end{enumerate}
A manifold $P$ endowed with a Poisson bracket on ${\mathcal F}(P)$ is called a {\bf Poisson manifold}.}

{\bf Proposition}. {\em Let ${\varphi}_t$ be a flow on  a Poisson manifold $P$ and let $H$:
$P \rightarrow \mathbb{R}$ be a smooth function on $P$. Then
\begin{enumerate}
\item for any $F \in {\mathcal F}(U)$, $U$ open in $P$, the equation of motion reads
\begin{equation}
\frac{d}{dt}\left(F \circ {\varphi}_t\right)=\{F, H\}\circ {\varphi}_t=\{F\circ {\varphi}_t, H\},
\end{equation}
or, for short,
\begin{equation}
\dot{F}=\{F, H\},
\end{equation}
if and only if ${\varphi}_t$ is the flow of $X_H$ (the Hamiltonian vector field of $H$).
\item
If ${\varphi}_t$ is the flow of $X_H$, then $H \circ {\varphi}_t=H$.
\end{enumerate}}

Secondly, in order to connect the Poisson manifold with our problem
we consider a manifold $P$ with (phase space) coordinates
$\xi^a$, $a=1,2,\cdots,2n$, such as $\xi^a=(x_k, x_l, x_{\gamma}; p_k, p_l, p_{\gamma})$ for our case,
and define as above the Poisson bracket
for arbitrary functions $F(\xi^a)$, $G(\xi^a)$, and $H(\xi^a) \in {\mathcal F}(P)$.
It can be proved\footnote{For details cf. Chapter 10 of Ref.~\cite{s16}.}
that if there exists a Lie-algebraic structure on a manifold,
such as the noncommutative spaces, eqs.~(8) and (9), the following
defines a Poisson manifold:
\begin{equation}
\{F,G\}=\{{\xi}^a,{\xi}^b\}\frac{{\partial F}}{{\partial {\xi}^a}}\frac{{\partial G}}{{\partial {\xi}^b}}.
\end{equation}
As a result, the Hamilton equation for any $F(\xi^a) \in {\mathcal F}(U)$, $U$ open in $P$,
takes the same form as eq.~(12). Specifically, the equation of motion thus reads
\begin{equation}
\dot{\xi^a}=\left\{\xi^a,H\right\},
\end{equation}
where $H=H(\xi^a)$ is Hamiltonian.

For the system of a nonrelativistic particle interacting with a constant external force,
$\vec{F}=(F_k, F_l, F_{\gamma})$,
its Hamiltonian takes the form,\footnote{The square of momenta appears in the Hamiltonian, that is the reason why we
have imposed eq.~(4), the vanishing Poisson brackets of momenta upon the phase space in order to avoid star-products in the following Hamilton
equation and Newton equation.}
\begin{equation}
H=\frac{p^2}{2m}+V(x),
\end{equation}
where $m$ is the mass of the particle and the potential is a linear function of spatial coordinates,
\begin{equation}
V(x)=-\sum_{a=k,l,{\gamma}} F_ax_a;\qquad F_a=-\frac{\partial V}{\partial{x_a}}= \textrm{Const.}
\end{equation}
At present we are ready to make a detailed Hamiltonian analysis for the system
on the noncommutative spaces Type I and Type II.

\subsection{Classical mechanics on Type I space}
On this noncommutative space, we at first derive the Hamilton equation in accordance with eq.~(8) and eqs.~(13)-(16),
\begin{eqnarray}
\dot{x}_k & = & \frac{p_k}{m}  -\frac{F_l-F_{\gamma }}{\kappa }t -\frac{F_{\gamma}}{\tilde{\kappa}} x_l,\\
\dot{x}_l & = & \frac{p_l}{m}  + \frac{F_k-F_{\gamma }}{\kappa } t +\frac{F_{\gamma}}{\tilde{\kappa}} x_k,\\
\dot{x}_{\gamma} & = &
\frac{p_{\gamma}}{m}-\frac{F_k-F_l}{\kappa}t+\frac{F_kx_l-F_lx_k}{\tilde{\kappa}};\\
\dot{p}_k & = & F_k-\frac{F_{\gamma}}{\tilde{\kappa}} p_l,\\
\dot{p}_l & = & F_l+\frac{F_{\gamma}}{\tilde{\kappa}} p_k,\\
\dot{p}_{\gamma} & = & F_{\gamma},
\end{eqnarray}
and then deduce the Newton equation by eliminating the momenta,
\begin{eqnarray}
m\ddot{x}_k & = & F_k -
m\frac{F_l-F_{\gamma}}{\kappa}+m\frac{(F_k-F_{\gamma})F_{\gamma}}{\kappa{\tilde{\kappa}}}t
+m\left(\frac{F_{\gamma}}{{\tilde{\kappa}}}\right)^2x_k-2m\frac{F_{\gamma}}{\tilde{\kappa}} \dot{x}_l,\\
m\ddot{x}_l & = & F_l +
m\frac{F_k-F_{\gamma}}{\kappa}+m\frac{(F_l-F_{\gamma})F_{\gamma}}{\kappa{\tilde{\kappa}}}t
+m\left(\frac{F_{\gamma}}{{\tilde{\kappa}}}\right)^2x_l+2m\frac{F_{\gamma}}{\tilde{\kappa}} \dot{x}_k,\\
m\ddot{x}_{\gamma} & = &
F_{\gamma}-m\frac{F_k-F_l}{\kappa}+m\frac{F_k\dot{x}_l-F_l\dot{x}_k}{\tilde{\kappa}}.
\end{eqnarray}
We notice that the noncommutativity brings about the appearance of extra forces which,
besides a constant contribution in each direction (the second term on the right side of each of the above equations),
are $t$-, $x$-, and $\dot{x}$-dependent, respectively.
The extra forces exist when the noncommutative parameters are finite, but vanish when these parameters tend to infinity.
This shows the consistency of our noncommutative generalization.
In particular, those extra forces related to two
noncommutative parameters or the square of one parameter,
such as the third and fourth terms in $\ddot{x}_k$
and $\ddot{x}_l$, exhibit entangled contributions from the noncommutativity {\em both} between different spatial
coordinates {\em and} between spatial coordinates and momenta.
The extra forces give rise to the deformation of the ordinary trajectories related to the Euclidean (commutative) space,
which can be seen obviously from the following solutions of the Newton equation,
\begin{eqnarray}
x_k(t) & = & -\frac{1}{m}\left(\frac{\tilde{\kappa }}{F_{\gamma }}\right)^2
\left[F_l+m\left(-\frac{F_k-F_{\gamma}}{\kappa}
+\left(\frac{F_{\gamma}}{\tilde{\kappa }}\right)^2x_{l0}\right)\right]
\sin\left(\frac{ F_{\gamma }}{\tilde{\kappa }}t\right) \nonumber\\
&&+\frac{1}{m}\left(\frac{\tilde{\kappa }}{F_{\gamma }}\right)^2\left[F_k+m \left(\frac{F_l-F_{\gamma }}{\kappa }
+\left(\frac{F_{\gamma}}{\tilde{\kappa}}\right)^2x_{k0}\right)\right]
\cos\left(\frac{ F_{\gamma }}{\tilde{\kappa }}t\right)\nonumber\\
&&+\frac{F_{\gamma}}{\tilde{\kappa}}\left[x_{k0}-\frac{\tilde{\kappa}}{F_{\gamma}}v_{l0}
+ \left(\frac{\tilde{\kappa}}{F_{\gamma}}\right)^2\frac{F_k}{m}\right]t
\sin \left(\frac{F_{\gamma}}{\tilde{\kappa}}t\right)\nonumber\\
&&+\frac{F_{\gamma}}{\tilde{\kappa}}\left[x_{l0}+\frac{\tilde{\kappa}}{F_{\gamma}}v_{k0}
+ \left(\frac{\tilde{\kappa}}{F_{\gamma}}\right)^2\frac{F_l}{m}\right]t
\cos \left(\frac{F_{\gamma}}{\tilde{\kappa}}t\right) \nonumber\\
&&- \frac{\tilde{\kappa}}{F_{\gamma }}\frac{F_k-F_{\gamma }}{\kappa}t
-\frac{1}{m}\left(\frac{\tilde{\kappa }}{F_{\gamma }}\right)^2\left(F_k+m \frac{F_l-F_{\gamma}}{\kappa}\right), \\
x_l(t) & = &\frac{1}{m}\left(\frac{\tilde{\kappa }}{F_{\gamma }}\right)^2
\left[F_k+m\left(\frac{F_l-F_{\gamma}}{\kappa}
+\left(\frac{F_{\gamma}}{\tilde{\kappa }}\right)^2x_{k0}\right)\right]
\sin\left(\frac{ F_{\gamma }}{\tilde{\kappa }}t\right) \nonumber\\
&&+\frac{1}{m}\left(\frac{\tilde{\kappa}}{F_{\gamma }}\right)^2\left[F_l+m \left(-\frac{F_k-F_{\gamma}}{\kappa}
+\left(\frac{F_{\gamma}}{\tilde{\kappa}}\right)^2x_{l0}\right)\right]
\cos\left(\frac{ F_{\gamma }}{\tilde{\kappa }}t\right)\nonumber\\
&&+\frac{F_{\gamma}}{\tilde{\kappa}}\left[x_{l0}+\frac{\tilde{\kappa}}{F_{\gamma}}v_{k0}
+ \left(\frac{\tilde{\kappa}}{F_{\gamma}}\right)^2\frac{F_l}{m}\right]t
\sin \left(\frac{F_{\gamma}}{\tilde{\kappa}}t\right) \nonumber\\
&& -\frac{F_{\gamma}}{\tilde{\kappa}}\left[x_{k0}-\frac{\tilde{\kappa}}{F_{\gamma}}v_{l0}
+ \left(\frac{\tilde{\kappa}}{F_{\gamma}}\right)^2\frac{F_k}{m}\right]t
\cos \left(\frac{F_{\gamma}}{\tilde{\kappa}}t\right) \nonumber\\
&&- \frac{\tilde{\kappa}}{F_{\gamma }}\frac{F_l-F_{\gamma }}{\kappa}t
-\frac{1}{m}\left(\frac{\tilde{\kappa }}{F_{\gamma }}\right)^2\left(F_l-m \frac{F_k-F_{\gamma}}{\kappa}\right),\\
x_{\gamma}(t)& = &
x_{\gamma0}+\left(v_{\gamma0}-\frac{F_kx_{l0}-F_lx_{k0}}{\tilde{\kappa}}\right)t+\left(\frac{F_{\gamma}}{m}
-\frac{F_k-F_l}{\kappa}\right)\frac{t^2}{2}\nonumber\\
&& +\frac{1}{\tilde{\kappa}}\int^t_0\left(F_kx_l(z)-F_lx_k(z)\right)dz,
\end{eqnarray}
where $x_{a0}$ and $v_{a0}$ with $a=k, l, \gamma$ stand for initial positions and velocities, respectively.

Let us analyze briefly the feature of this kind of solutions of the Newton equation.
In $k$- and $l$-directions the solutions
depend on in general three functions: a periodic function with the period $2\pi\tilde{\kappa}/F_{\gamma}$,
time variable times a periodic function
with the same period but a different phase, and a linear function of time. In $\gamma$-direction the solution
relates to a linear and quadratic functions of time, and to an integration of the solutions in the other
two directions. Therefore this kind of solutions deforms greatly the particle's trajectories we are familiar with
on the Euclidean (commutative) space.
Incidentally, for the special case of no external forces, {\em i.e.} $\vec{F}=0$, the Newton equation
(eqs.~(23)-(25)) becomes the same form as that on the Euclidean space.
This implies that the noncommutativity of Type I space is not
active for a free particle, rather than that Type I space is commutative and isotropic,
which can be seen clearly from the
case of a nonzero external force on this space and from the cases of both zero and nonzero external forces
on Type II space.

In order to plot trajectories, we simplify, without any loss of generality, the solutions (eqs.~(26)-(28)) by
letting $x_{a0}$, $v_{k0}$, $v_{{\gamma}0}$, $F_k$, and $F_l$ be zero, while maintaining $v_{l0}$,
$F_{\gamma}$, $m$, $\kappa$, and
$\tilde{\kappa}$ nonzero. As a result, we arrive at the simpler formulae of the solutions,
\begin{eqnarray}
x_k(t) & = & -\frac{{\tilde{\kappa}}^2}{{\kappa}F_{\gamma}}\sin\left(\frac{F_{\gamma}}{\tilde{\kappa}}t\right)
+\frac{{\tilde{\kappa}}^2}{{\kappa}F_{\gamma}}\left[1-\cos\left(\frac{F_{\gamma}}{\tilde{\kappa}}t\right)\right]
-v_{l0}t\sin\left(\frac{F_{\gamma}}{\tilde{\kappa}}t\right)+\frac{{\tilde{\kappa}}}{{\kappa}}t,\\
x_l(t) & = &-\frac{{\tilde{\kappa}}^2}{{\kappa}F_{\gamma}}\sin\left(\frac{F_{\gamma}}{\tilde{\kappa}}t\right)
-\frac{{\tilde{\kappa}}^2}{{\kappa}F_{\gamma}}\left[1-\cos\left(\frac{F_{\gamma}}{\tilde{\kappa}}t\right)\right]
+v_{l0}t\cos\left(\frac{F_{\gamma}}{\tilde{\kappa}}t\right)+\frac{{\tilde{\kappa}}}{{\kappa}}t,\\
x_{\gamma}(t)& = & \frac{F_{\gamma}}{2m}t^2.
\end{eqnarray}
The corresponding trajectories are illustrated in Figure 1 for typical values of the nonzero parameters, $v_{l0}$,
$F_{\gamma}$, $m$, $\kappa$, and $\tilde{\kappa}$.
\begin{figure}
\centering
\hfill
\includegraphics[width=0.32\textwidth]{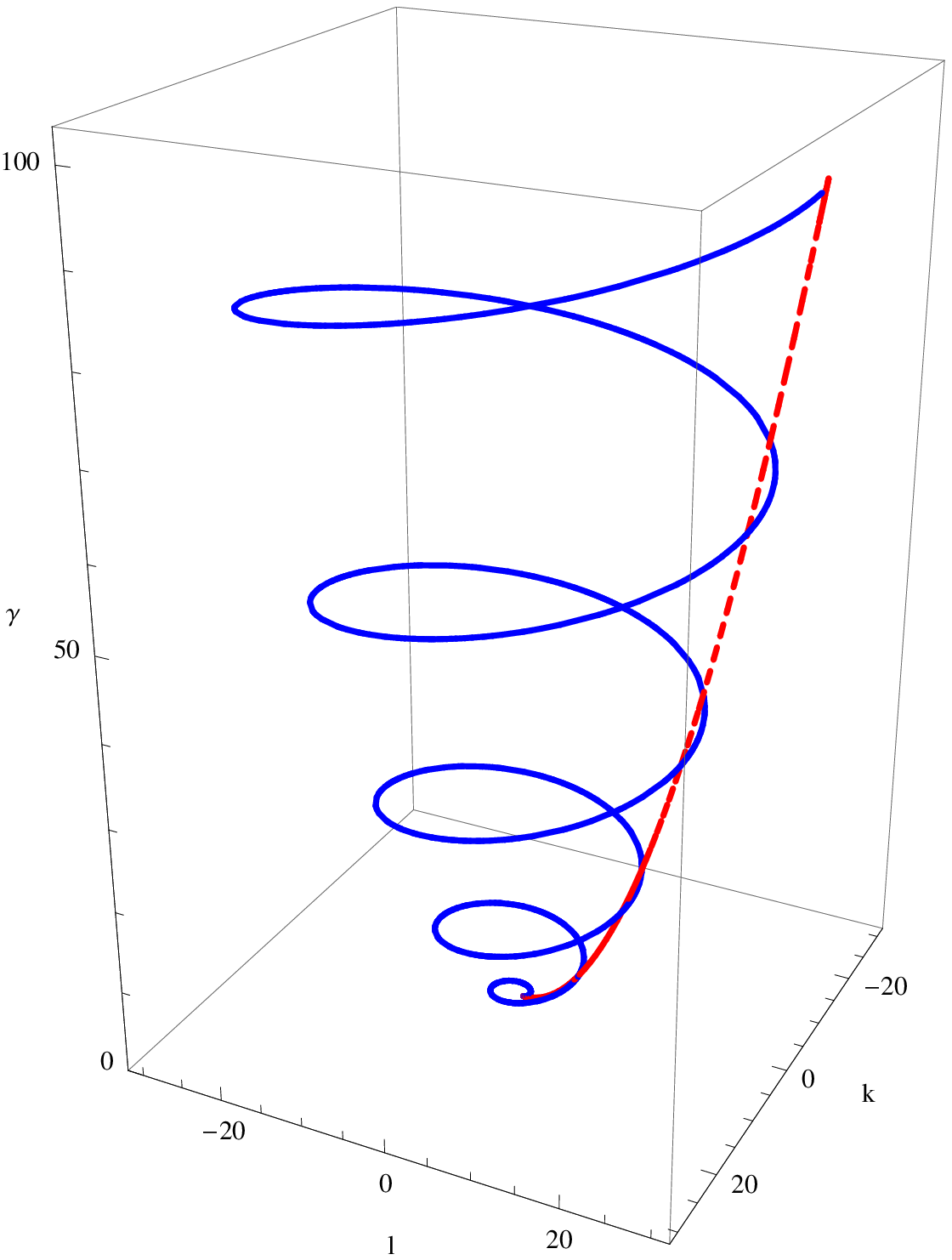}
\hfill
\includegraphics[width=0.32\textwidth]{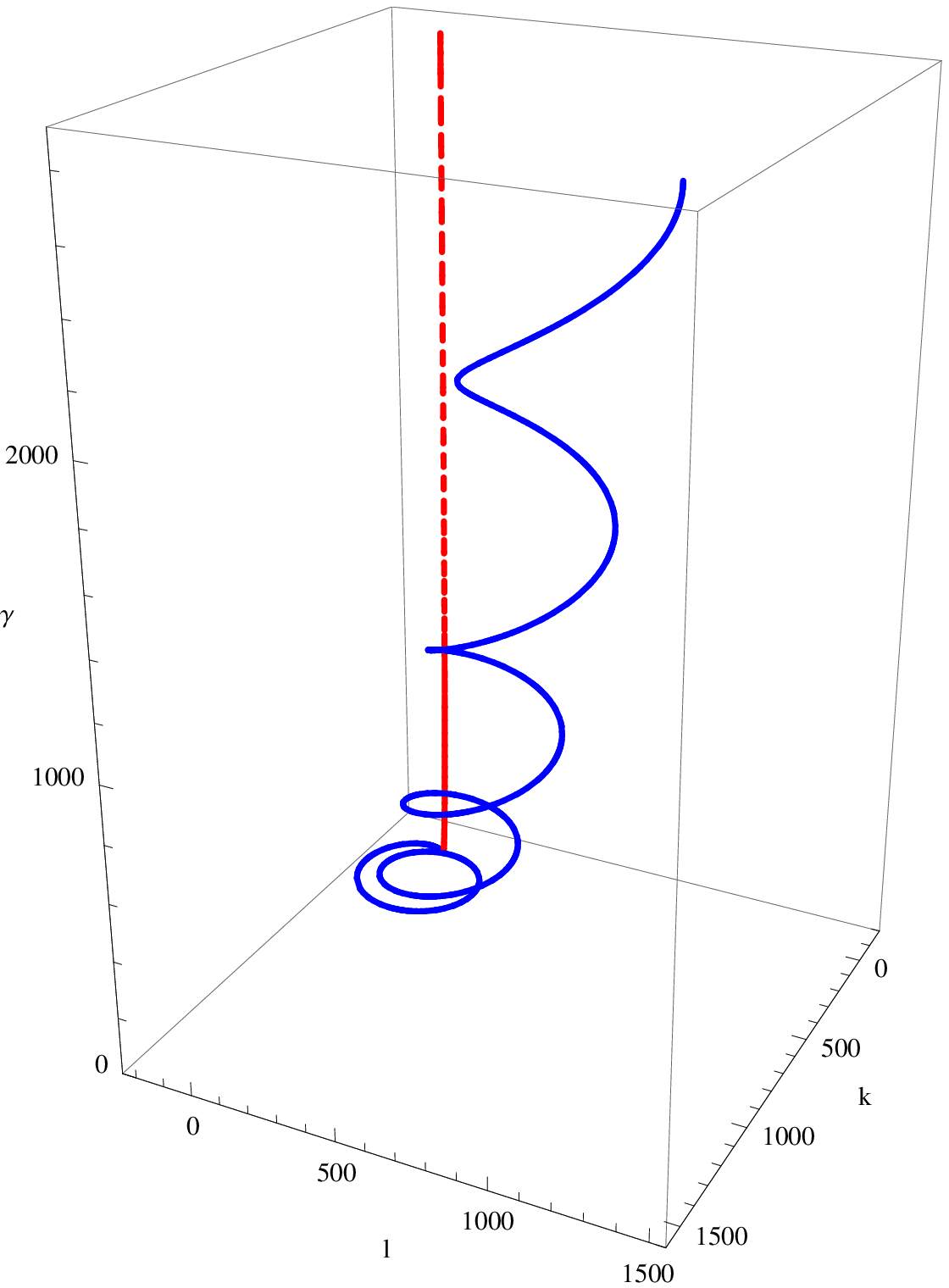}
\hspace*{\fill}
\caption{The left and right figures correspond to different noncommutative parameters
$(\kappa, \tilde{\kappa})=(410, 10)$ and $(10, 410)$, and different $\gamma$-components of the external force
$F_\gamma=\tilde{\kappa}=(10, 410)$, respectively,
but to same values of the rest of parameters,
{\em i.e.} $v_{l0}=1$ and $m=1$. Note that the $\gamma$-directions of the left and right figures
are compressed by 50 and 70 times, respectively, in order to reflect the characteristic of the trajectories.
The time variable runs from 0 to $10\pi\tilde{\kappa}/F_{\gamma}$.
To make a comparison between the deformed and undeformed
cases, the latter $(\kappa=\infty, \tilde{\kappa}=\infty)$ is plotted by a dash line.}
\end{figure}

\subsection{Classical mechanics on Type II space}
On this noncommutative space, we write the Hamilton equation in accordance with eq.~(9) and eqs.~(13)-(16) as follows:
\begin{eqnarray}
\dot{x}_k & = & \frac{p_k}{m}+\frac{F_{\gamma}}{\kappa} t-\frac{F_{\gamma}}{\tilde{\kappa}} x_l,\\
\dot{x}_l & = & \frac{p_l}{m}-\frac{F_{\gamma}}{\kappa} t+\frac{F_{\gamma}}{\tilde{\kappa}} x_k,\\
\dot{x}_{\gamma} & = &
\frac{p_{\gamma}}{m}-\frac{p_kx_l+p_lx_k}{m\bar{\kappa}}-\frac{F_k-F_l}{\kappa}t+\frac{F_kx_l-F_lx_k}{\tilde{\kappa}};\\
\dot{p}_k & = & F_k-\frac{F_{\gamma}}{\bar{\kappa}} x_l-\frac{F_{\gamma}}{\tilde{\kappa}} p_l,\\
\dot{p}_l & = & F_l-\frac{F_{\gamma}}{\bar{\kappa}} x_k+\frac{F_{\gamma}}{\tilde{\kappa}} p_k,\\
\dot{p}_{\gamma} & = & F_{\gamma}.
\end{eqnarray}
Note that the bilinear terms, $p_kx_l$ and $p_lx_k$ in $\dot{x}_{\gamma}$, do not cause any ambiguity because
the Poisson brackets of $p_k$ and $x_l$ and of $p_l$ and $x_k$ are vanishing, see eq.~(9).
This property guarantees the consistency of the Hamilton equation, that is, the Hamilton equation is not involved in
the ordering of coordinates and momenta, which coincides with our starting point that no star-products are needed.
Correspondingly, we derive the Newton equation by eliminating the momenta,
\begin{eqnarray}
m\ddot{x}_k & = & F_k + m\frac{F_{\gamma}}{\kappa}-m\frac{{F_{\gamma}}^2}{\kappa{\tilde{\kappa}}}t
+m\left(\frac{F_{\gamma}}{{\tilde{\kappa}}}\right)^2x_k-\frac{F_{\gamma}}{\bar{\kappa}}x_l
-2m\frac{F_{\gamma}}{\tilde{\kappa}} \dot{x}_l,\\
m\ddot{x}_l & = & F_l - m\frac{F_{\gamma}}{\kappa}-m\frac{{F_{\gamma}}^2}{\kappa{\tilde{\kappa}}}t
+m\left(\frac{F_{\gamma}}{{\tilde{\kappa}}}\right)^2x_l-\frac{F_{\gamma}}{\bar{\kappa}}x_k
+2m\frac{F_{\gamma}}{\tilde{\kappa}} \dot{x}_k,\\
m\ddot{x}_{\gamma} & = & F_{\gamma}-m\frac{F_k-F_l}{\kappa}-\frac{m}{\bar{\kappa}}\ddot{\left(x_kx_l\right)}
-m\frac{F_{\gamma}}{{\kappa}\bar{\kappa}}\left(x_k-x_l+t\dot{x}_k-t\dot{x}_l\right)\nonumber\\
& &
+m\frac{F_k\dot{x}_l-F_l\dot{x}_k}{\tilde{\kappa}}
+m\frac{F_{\gamma}}{\tilde{\kappa}\bar{\kappa}}\left(\dot{\left({x_k}{x_k}\right)}-\dot{\left({x_l}{x_l}\right)}\right).
\end{eqnarray}
For the special case of no external forces, {\em i.e.} $\vec{F}=0$, the above equations take a quite simple
form, $m\ddot{x}_k  = 0$, $m\ddot{x}_l  = 0$,
and $m\ddot{x}_{\gamma}=-{m}\ddot{\left(x_kx_l\right)}/{\bar{\kappa}}=-2{m}v_{k0}v_{l0}/{\bar{\kappa}}$. This
comes to the conclusion that the noncommutativity of Type II space (only the finite ${\bar{\kappa}}$)
is active even for a free particle, which is different from the case of Type I space.

Let us have a careful look at the extra forces in the equations of motion (eqs.~(38)-(40)).
In general, besides the constant external force $\vec{F}=(F_k, F_l, F_{\gamma})$,
five extra forces emerge from the noncommutativity of the phase space in $k$- and $l$-directions, and
more extra forces with complicated formulations appear in ${\gamma}$-direction.
In $k$- and $l$-directions the first extra force is constant, the second,
the third and fourth, and the last are proportional to time, spatial coordinate, and velocity, respectively.
In ${\gamma}$-direction, moreover, the noncommutativity gives rise to various formulations of extra forces in which
the different ones from that of $k$- and $l$-directions are those proportional to
$\ddot{\left(x_kx_l\right)}$, $\left(t\dot{x}_k-t\dot{x}_l\right)$, and
$\left(\dot{\left({x_k}{x_k}\right)}-\dot{\left({x_l}{x_l}\right)}\right)$, respectively.
We mention that the bilinear term $x_kx_l$ does not cause any ordering ambiguity because of the vanishing
Poisson bracket $\{x_k,x_{l}\}=0$, see eq.~(9). This guarantees the consistency of the Newton equation in
${\gamma}$-direction, which reflects the independence of star-products.
Note also that some extra forces in the Newton equation
are related to two noncommutative parameters or the square of one parameter,
such as the third and fourth terms in $\ddot{x}_k$ and $\ddot{x}_l$,
and the fourth and sixth terms in $\ddot{x}_{\gamma}$,
which exhibits entangled contributions that arise from the noncommutativity {\em both} between different spatial
coordinates {\em and} between spatial coordinates and momenta.
We emphasize that all the extra forces vanish when the noncommutative parameters tend to infinity,
which makes our noncommutative generalization consistent.
Incidentally, the three new kinds of extra forces in $\ddot{x}_{\gamma}$, i.e. $\ddot{\left(x_kx_l\right)}$, $\left(t\dot{x}_k-t\dot{x}_l\right)$, and
$\left(\dot{\left({x_k}{x_k}\right)}-\dot{\left({x_l}{x_l}\right)}\right)$, do not appear in the equations of motion
in Ref.~\cite{s5}.

From the Newton equation we find that the solution in ${\gamma}$-direction can be expressed by the
solutions in the other two directions as follows:
\begin{eqnarray}
x_{\gamma}(t)& = & x_{\gamma0}+Ct+\left(\frac{F_{\gamma}}{m}-\frac{F_k-F_l}{\kappa}\right)\frac{t^2}{2}
-\frac{x_k(t)x_l(t)-x_{k0}x_{l0}}{\bar{\kappa}}\nonumber\\
& & +\int^t_0\left[\frac{F_kx_l(z)-F_lx_k(z)}{\tilde{\kappa}}-\frac{F_{\gamma}}{\bar{\kappa}}
\left[\frac{z\left(x_k(z)-x_l(z)\right)}{\kappa}
-\frac{x_k^2(z)-x_l^2(z)}{\tilde{\kappa}}\right]\right]dz,
\end{eqnarray}
where $C$ is defined by
\begin{equation}
C \equiv v_{\gamma0}+\frac{x_{k0}v_{l0}+x_{l0}v_{k0}}{\bar{\kappa}}
-\frac{F_kx_{l0}-F_lx_{k0}}{\tilde{\kappa}}
-\frac{F_{\gamma}}{\tilde{\kappa}{\bar{\kappa}}}\left(x_{k0}^2-x_{l0}^2\right),
\end{equation}
and $x_{a0}$ and $v_{a0}$ with $a=k, l, \gamma$ denote initial positions and velocities, respectively.
As to the solutions in $k$- and $l$-directions, we have to discuss in category which depends on
factor $\varepsilon$. This factor is composed of the
${\gamma}$-component of the external force, the mass, and two of the three noncommutative parameters as follows:
\begin{equation}
\varepsilon \equiv \frac{F_{\gamma}}{{\tilde{\kappa}}^2/(m\bar{\kappa})}.
\end{equation}
It is dimensionless.

\subsubsection{Case (i): $\varepsilon < 1$}
For this case we solve the Newton equation in $k$- and $l$-directions and obtain the solutions,
\begin{eqnarray}
x_k(t) & = &\left[-\frac{F_{\gamma}}{\tilde{\kappa}}\left(\left(x_{k0}-B_k\right)
-\varepsilon\left(x_{l0}-B_l\right)\right)
+\frac{1+2\varepsilon}{2}\left(v_{k0}-A\right)
+\frac{v_{l0}-A}{2}\right]\frac{1}{\omega_1}\sin\omega_1t \nonumber \\
& &
+\left[\frac{1-2\varepsilon}{2}\left(x_{k0}-B_k\right)+\frac{x_{l0}-B_l}{2}
+\frac{\varepsilon{\tilde{\kappa}}}{F_{\gamma}}\left(v_{l0}-A\right)\right]\cos\omega_1t \nonumber \\
& &
+\left[\frac{F_{\gamma}}{\tilde{\kappa}}\left(\left(x_{k0}-B_k\right)
-\varepsilon\left(x_{l0}-B_l\right)\right)
+\frac{1-2\varepsilon}{2}\left(v_{k0}-A\right)-\frac{v_{l0}-A}{2}
\right]\frac{1}{\omega_2}\sinh\omega_2t \nonumber\\
& &
+\left[\frac{1+2\varepsilon}{2}\left(x_{k0}-B_k\right)-\frac{x_{l0}-B_l}{2}
-\frac{\varepsilon{\tilde{\kappa}}}{F_{\gamma}}\left(v_{l0}-A\right)\right]\cosh\omega_2t \nonumber\\
& & +At+B_k,
\end{eqnarray}
and
\begin{eqnarray}
x_l(t) & = &\left[\frac{F_{\gamma}}{\tilde{\kappa}}\left(\left(x_{l0}-B_l\right)
-\varepsilon\left(x_{k0}-B_k\right)\right)
+\frac{1+2\varepsilon}{2}\left(v_{l0}-A\right)+\frac{v_{k0}-A}{2}\right]\frac{1}{\omega_1}\sin\omega_1t \nonumber\\
& &
+\left[\frac{1-2\varepsilon}{2}\left(x_{l0}-B_l\right)
+\frac{x_{k0}-B_k}{2}-\frac{\varepsilon{\tilde{\kappa}}}{F_{\gamma}}\left(v_{k0}-A\right)\right]\cos\omega_1t
\nonumber\\
& &
+\left[-\frac{F_{\gamma}}{\tilde{\kappa}}\left(\left(x_{l0}-B_l\right)
-\varepsilon\left(x_{k0}-B_k\right)\right)
+\frac{1-2\varepsilon}{2}\left(v_{l0}-A\right)-\frac{v_{k0}-A}{2}\right]\frac{1}{\omega_2}\sinh\omega_2t\nonumber\\
& &
+\left[\frac{1+2\varepsilon}{2}\left(x_{l0}-B_l\right)-\frac{x_{k0}-B_k}{2}
+\frac{\varepsilon{\tilde{\kappa}}}{F_{\gamma}}\left(v_{k0}-A\right)\right]\cosh\omega_2t
\nonumber\\
& & +At+B_l,
\end{eqnarray}
where $A$, $B_k$, $B_l$, $\omega_1$, and $\omega_2$ are independent of time and defined by
\begin{eqnarray}
A&\equiv&-\frac{\varepsilon}{1-\varepsilon}\frac{\tilde{\kappa}}{\kappa}, \\
B_k&\equiv&\frac{1}{F_{\gamma}(1-\varepsilon)}\left[\frac{\bar{\kappa}(\varepsilon{F_k}+{F_l})}{1+\varepsilon}
-\frac{\varepsilon{\tilde{\kappa}}^2}{\kappa}\right], \\
B_l&\equiv&\frac{1}{F_{\gamma}(1-\varepsilon)}\left[\frac{\bar{\kappa}(F_k+\varepsilon{F_l})}{1+\varepsilon}
+\frac{\varepsilon{\tilde{\kappa}}^2}{\kappa}\right], \\
\omega_1&\equiv&\frac{F_{\gamma}}{\tilde{\kappa}}\sqrt{\frac{1+\varepsilon}{\varepsilon}}, \\
\omega_2&\equiv&\frac{F_{\gamma}}{\tilde{\kappa}}\sqrt{\frac{1-\varepsilon}{\varepsilon}}.
\end{eqnarray}
The characteristic of the solutions is the linear combination of trigonometric functions and hyperbolic functions.

In order to plot trajectories, we simplify, as we did in subsection 3.1, the solutions  by
letting $x_{a0}$, $v_{k0}$, $v_{{\gamma}0}$, $F_k$, and $F_l$ be zero, while maintaining $v_{l0}$,
$F_{\gamma}$, $m$, $\kappa$,
$\tilde{\kappa}$, and $\bar{\kappa}$ nonzero. Therefore, the solutions reduce to
\begin{eqnarray}
x_k(t) & = & \frac{v_{l0}}{2}\frac{\sin{\omega}_1t}{{\omega}_1}
+\frac{m\bar{\kappa}}{\tilde{\kappa}}v_{l0}\cos{\omega_1}t
+ \left(\frac{\varepsilon}{1-\varepsilon}\frac{\tilde{\kappa}}{\kappa}
-\frac{v_{l0}}{2}\right)\frac{\sinh{\omega}_2t}{{\omega}_2} \nonumber \\
& &
+\left(\frac{1}{1-\varepsilon}\frac{m\bar{\kappa}}{\kappa}
-\frac{m\bar{\kappa}}{\tilde{\kappa}}v_{l0}\right)\cosh{\omega}_2t
-\frac{\varepsilon}{1-\varepsilon}\frac{\tilde{\kappa}}{\kappa}\,t
-\frac{1}{1-\varepsilon}\frac{m\bar{\kappa}}{\kappa},\\
x_l(t) & = & \frac{1+2\varepsilon}{2}v_{l0}\frac{\sin{\omega}_1t}{{\omega}_1}
+\left(\frac{\varepsilon}{1-\varepsilon}\frac{\tilde{\kappa}}{\kappa}
+\frac{1-2\varepsilon}{2}v_{l0}\right)\frac{\sinh{\omega}_2t}{{\omega}_2}
\nonumber \\
& &-\frac{1}{1-\varepsilon}\frac{m\bar{\kappa}}{\kappa}\cosh{\omega}_2t
-\frac{\varepsilon}{1-\varepsilon}\frac{\tilde{\kappa}}{\kappa}\,t
+\frac{1}{1-\varepsilon}\frac{m\bar{\kappa}}{\kappa},\\
x_{\gamma}(t) & = &
\frac{F_{\gamma}}{2m}t^2-\frac{x_k(t)x_l(t)}{\bar{\kappa}} 
-\frac{F_{\gamma}}{\bar{\kappa}}\int^t_0\left[\frac{z\left(x_k(z)-x_l(z)\right)}{\kappa}
-\frac{x_k^2(z)-x_l^2(z)}{\tilde{\kappa}}\right]dz.
\end{eqnarray}
The corresponding trajectories are
illustrated in Figure 2 for typical values of the nonzero parameters, $v_{l0}$,
$F_{\gamma}$, $m$, $\kappa$, $\tilde{\kappa}$, and $\bar{\kappa}$ .
\begin{figure}
\centering
\hfill
\includegraphics[width=0.32\textwidth]{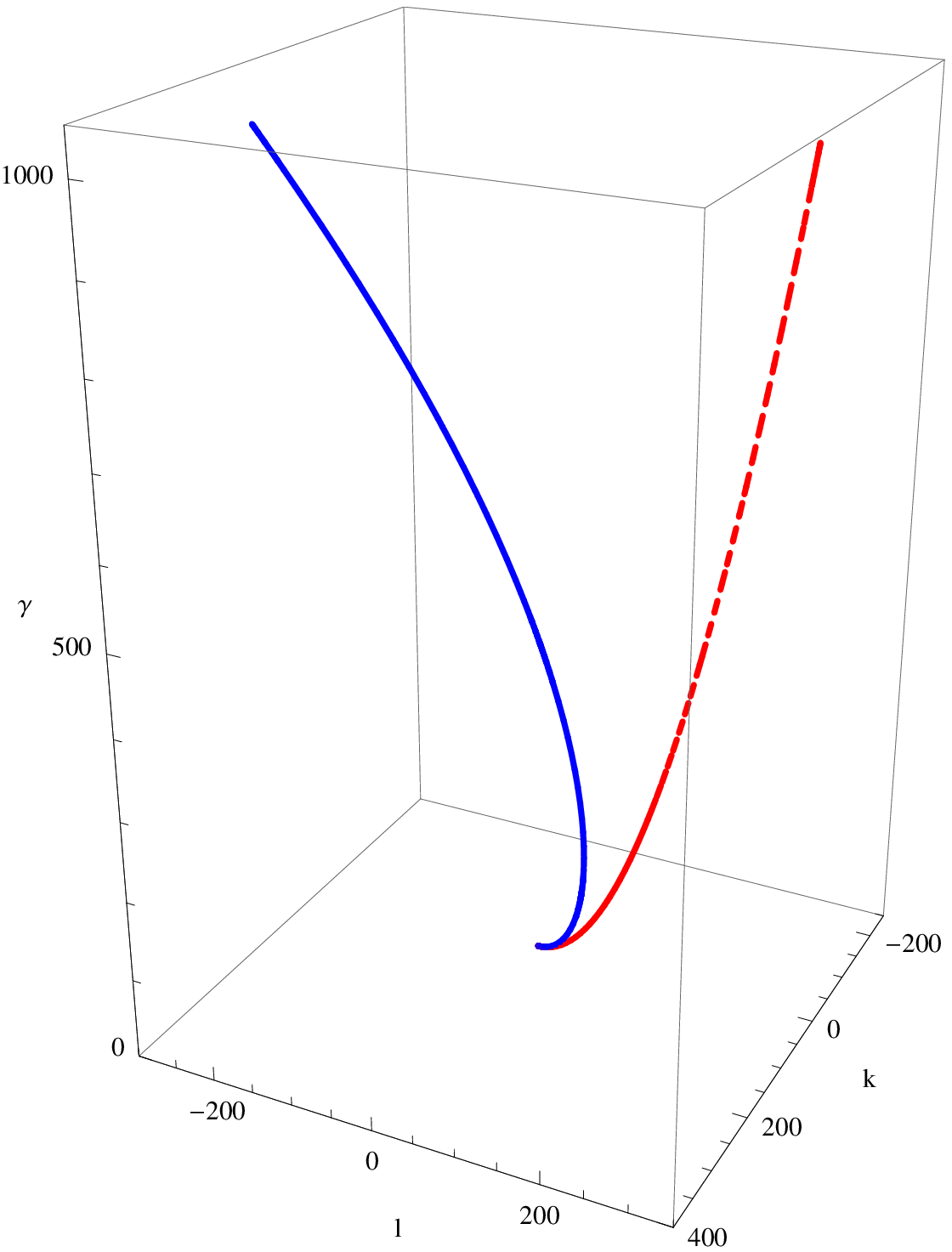}
\hfill
\includegraphics[width=0.32\textwidth]{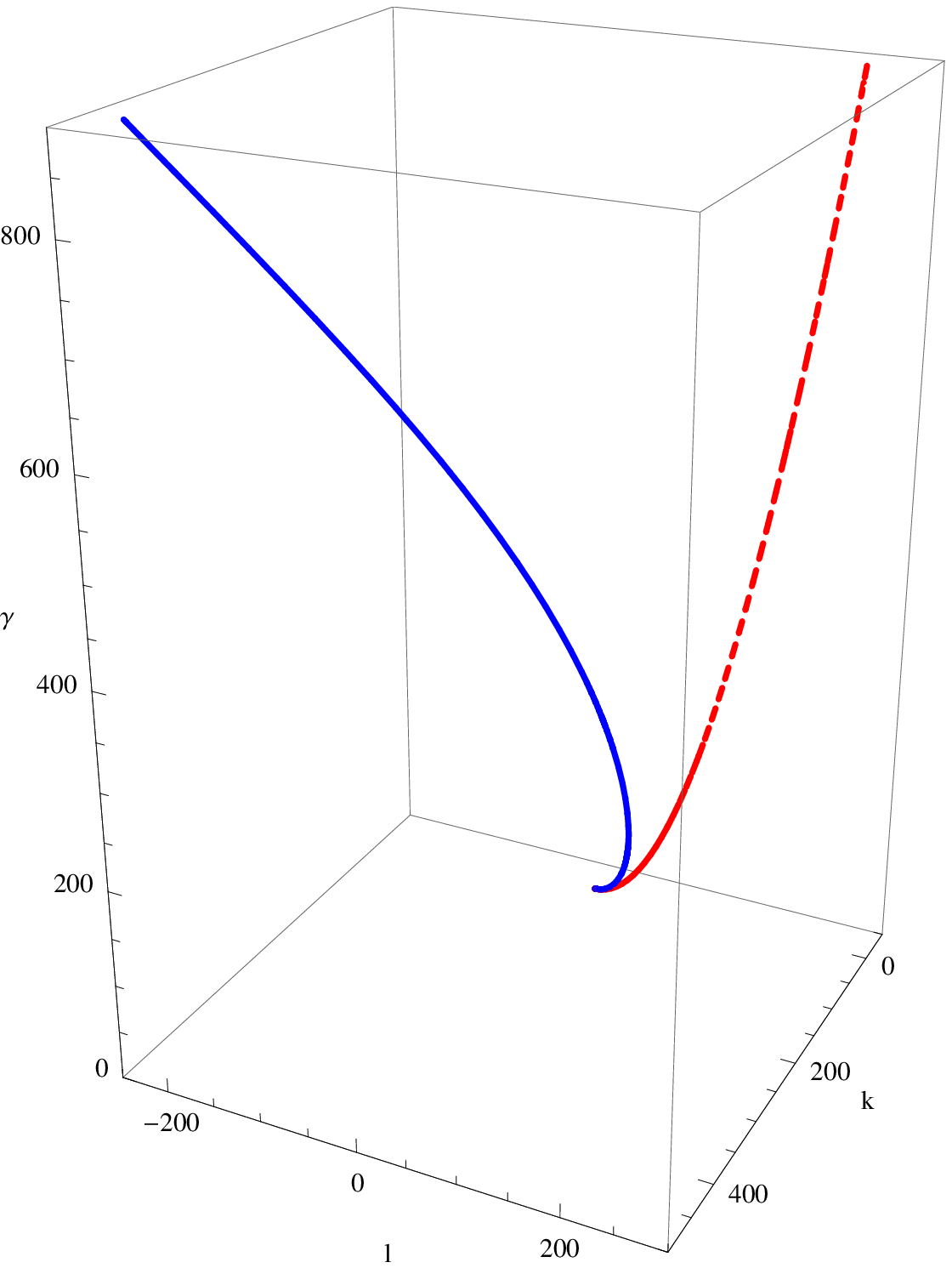}
\hspace*{\fill}
\caption{The left and right figures correspond to different $\gamma$-components of the external
force $F_\gamma=\mu{\tilde{\kappa}}^2/\bar{\kappa}$ with the coefficient $\mu$=0.4 and 0.6, respectively,
but to same values of the rest of parameters, {\em i.e.} $\kappa=10, \tilde{\kappa}=10, \bar{\kappa}=500$,
$v_{l0}=1$, and $m=1$. Note that the $\gamma$-directions of the left and right figures
are compressed by 3.7 and 4.1 times, respectively, in order to reflect the characteristic of the trajectories.
The time variable runs from 0 to
$6\pi/\sqrt{F_{\gamma}(1/\bar{\kappa}-F_{\gamma}/{\tilde{\kappa}}^2)}$.
To make a comparison between the deformed and undeformed
cases, the latter $(\kappa=\infty, \tilde{\kappa}=\infty, \bar{\kappa}=\infty)$ is plotted by a dash line.}
\end{figure}

\subsubsection{Case (ii): $\varepsilon > 1$}
For this case we get the solutions in $k$- and $l$-directions as follows:
\begin{eqnarray}
x_k(t) & = & \left[-\frac{F_{\gamma}}{\tilde{\kappa}}\left(\left(x_{k0}-B_k\right)
-\varepsilon\left(x_{l0}-B_l\right)\right)
+\frac{1+2\varepsilon}{2}\left(v_{k0}-A\right)
+\frac{v_{l0}-A}{2}\right]\frac{1}{\omega_1}\sin\omega_1t \nonumber\\
& &
+\left[\frac{1-2\varepsilon}{2}\left(x_{k0}-B_k\right)+\frac{x_{l0}-B_l}{2}
+\frac{\varepsilon{\tilde{\kappa}}}{F_{\gamma}}\left(v_{l0}-A\right)\right]\cos\omega_1t
\nonumber\\
& &
+\left[\frac{F_{\gamma}}{\tilde{\kappa}}\left(\left(x_{k0}-B_k\right)
-\varepsilon\left(x_{l0}-B_l\right)\right)
+\frac{1-2\varepsilon}{2}\left(v_{k0}-A\right)-\frac{v_{l0}-A}{2}
\right]\frac{1}{\omega_2^{\prime}}\sin\omega_2^{\prime}t \nonumber\\
& &
+\left[\frac{1+2\varepsilon}{2}\left(x_{k0}-B_k\right)-\frac{x_{l0}-B_l}{2}
-\frac{\varepsilon{\tilde{\kappa}}}{F_{\gamma}}\left(v_{l0}-A\right)\right]\cos\omega_2^{\prime}t
\nonumber\\
& & +At+B_k,
\end{eqnarray}
and
\begin{eqnarray}
x_l(t) & = &\left[\frac{F_{\gamma}}{\tilde{\kappa}}\left(\left(x_{l0}-B_l\right)
-\varepsilon\left(x_{k0}-B_k\right)\right)
+\frac{1+2\varepsilon}{2}\left(v_{l0}-A\right)+\frac{v_{k0}-A}{2}\right]\frac{1}{\omega_1}\sin\omega_1t
\nonumber\\
& &
+\left[\frac{1-2\varepsilon}{2}\left(x_{l0}-B_l\right)
+\frac{x_{k0}-B_k}{2}-\frac{\varepsilon{\tilde{\kappa}}}{F_{\gamma}}\left(v_{k0}-A\right)\right]\cos\omega_1t
\nonumber\\
& &
+\left[-\frac{F_{\gamma}}{\tilde{\kappa}}\left(\left(x_{l0}-B_l\right)
-\varepsilon\left(x_{k0}-B_k\right)\right)
+\frac{1-2\varepsilon}{2}\left(v_{l0}-A\right)-\frac{v_{k0}-A}{2}\right]
\frac{1}{\omega_2^{\prime}}\sin\omega_2^{\prime}t\nonumber\\
& &
+\left[\frac{1+2\varepsilon}{2}\left(x_{l0}-B_l\right)-\frac{x_{k0}-B_k}{2}
+\frac{\varepsilon{\tilde{\kappa}}}{F_{\gamma}}\left(v_{k0}-A\right)\right]\cos\omega_2^{\prime}t
\nonumber\\
& & +At+B_l,
\end{eqnarray}
where $\omega_2^{\prime}$, a real parameter, is defined by
\begin{equation}
\omega_2^{\prime}\equiv\frac{F_{\gamma}}{\tilde{\kappa}}\sqrt{\frac{\varepsilon-1}{\varepsilon}}.
\end{equation}
Note that in this case the solutions contain two periodic functions with periods  $2\pi/{\omega_1}$ and
$2\pi/{\omega}_2^{\prime}$, respectively.

As dealt with in the previous case, we still maintain $v_{l0}$, $F_{\gamma}$, $m$, $\kappa$,
$\tilde{\kappa}$, and $\bar{\kappa}$ nonzero, and reduce the solutions to be,
\begin{eqnarray}
x_k(t) & = & \frac{v_{l0}}{2}\frac{\sin{\omega}_1t}{{\omega}_1}
+\frac{m\bar{\kappa}}{\tilde{\kappa}}v_{l0}\cos{\omega_1}t
+ \left(\frac{\varepsilon}{1-\varepsilon}\frac{\tilde{\kappa}}{\kappa}
-\frac{v_{l0}}{2}\right)\frac{\sin{\omega}_2^{\prime}t}{{\omega}_2^{\prime}} \nonumber \\
& &
+\left(\frac{1}{1-\varepsilon}\frac{m\bar{\kappa}}{\kappa}
-\frac{m\bar{\kappa}}{\tilde{\kappa}}v_{l0}\right)\cos{\omega}_2^{\prime}t
-\frac{\varepsilon}{1-\varepsilon}\frac{\tilde{\kappa}}{\kappa}\,t
-\frac{1}{1-\varepsilon}\frac{m\bar{\kappa}}{\kappa},\\
x_l(t) & = & \frac{1+2\varepsilon}{2}v_{l0}\frac{\sin{\omega}_1t}{{\omega}_1}
+\left(\frac{\varepsilon}{1-\varepsilon}\frac{\tilde{\kappa}}{\kappa}
+\frac{1-2\varepsilon}{2}v_{l0}\right)\frac{\sin{\omega}_2^{\prime}t}{{\omega}_2^{\prime}}
\nonumber \\
& &-\frac{1}{1-\varepsilon}\frac{m\bar{\kappa}}{\kappa}\cos{\omega}_2^{\prime}t
-\frac{\varepsilon}{1-\varepsilon}\frac{\tilde{\kappa}}{\kappa}\,t
+\frac{1}{1-\varepsilon}\frac{m\bar{\kappa}}{\kappa},
\end{eqnarray}
where $x_{\gamma}(t)$ keeps the same form as that in the previous case but relates to $x_k(t)$ and $x_l(t)$
in this case.

The corresponding trajectories are
illustrated in Figure 3 for typical values of the nonzero parameters, $v_{l0}$, $F_{\gamma}$, $m$, $\kappa$,
$\tilde{\kappa}$, and $\bar{\kappa}$.
\begin{figure}
\centering
\hfill
\includegraphics[width=0.32\textwidth]{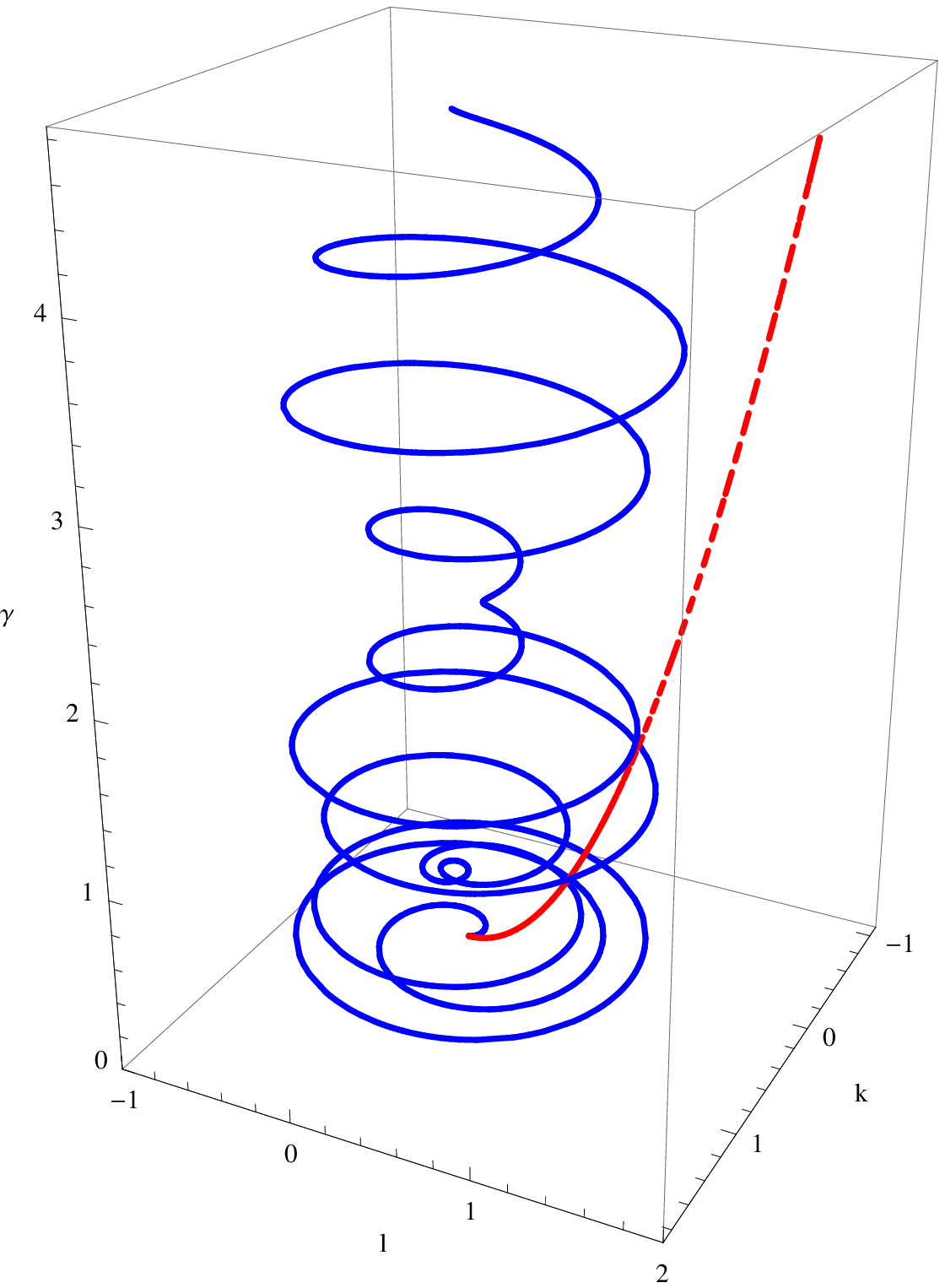}
\hfill
\includegraphics[width=0.32\textwidth]{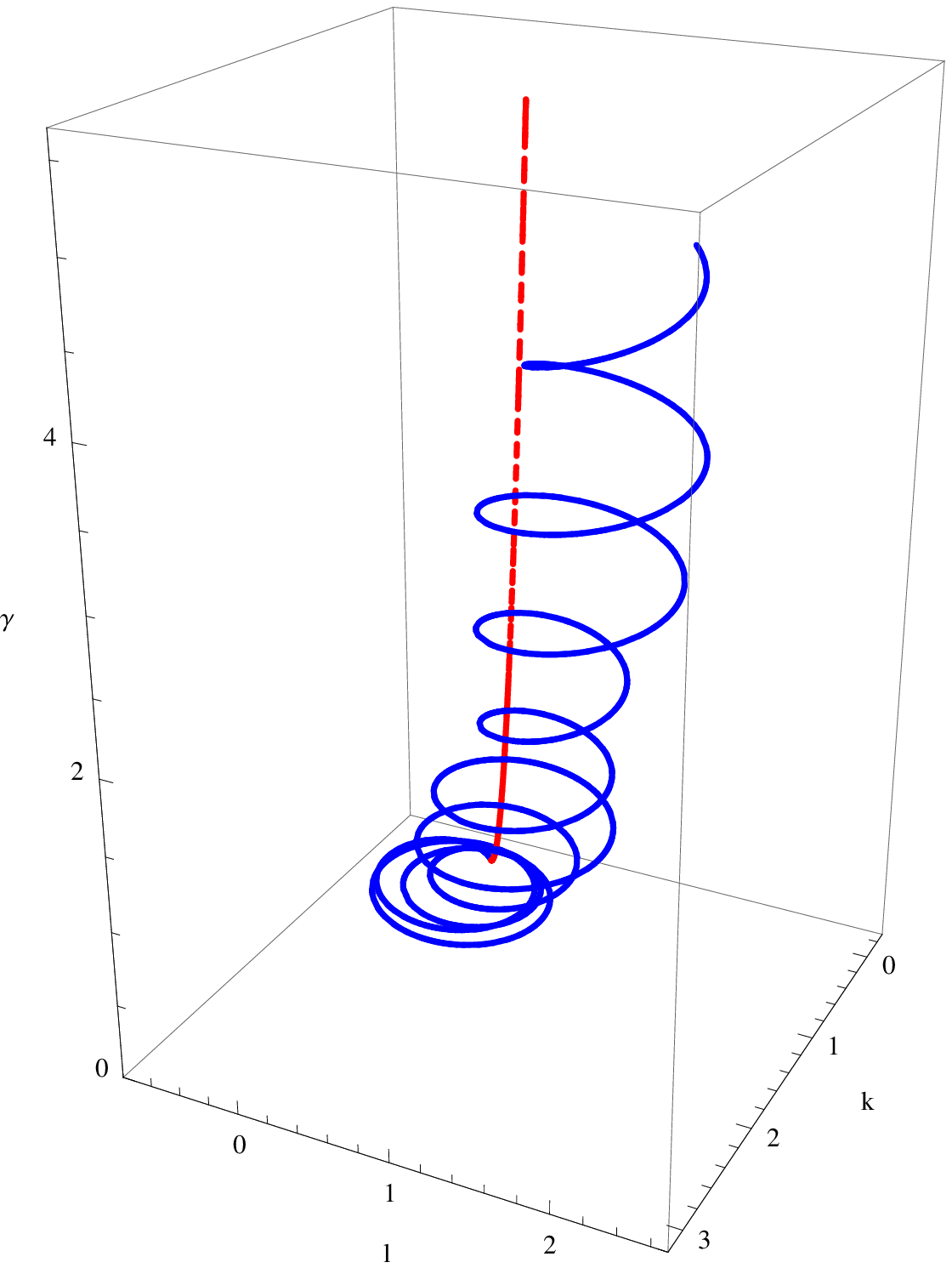}
\hspace*{\fill}
\caption{The left and right figures correspond to different noncommutative parameters
$(\kappa, \tilde{\kappa}, \bar{\kappa})=(700, 100, 10)$ and $(100, 700, 10)$,
and different $\gamma$-components of the external force
$F_\gamma=4\tilde{\kappa}^2/\bar{\kappa}=(4000, 196000)$, respectively,
but to same values of the rest of parameters,
{\em i.e.} $v_{l0}=1$ and $m=1$. Note that the $\gamma$-directions of the left and right figures
are compressed by 1400 and 1250 times, respectively, in order to reflect the characteristic of the trajectories.
The time variable runs from 0 to $10\pi\bar{\kappa}/(\sqrt{3}\tilde{\kappa})$.
To make a comparison between the deformed and undeformed
cases, the latter $(\kappa=\infty, \tilde{\kappa}=\infty, \bar{\kappa}=\infty)$ is plotted by a dash line.}
\end{figure}

\subsubsection{Case (iii): $\varepsilon = 1$}
For this case, we get the critical solutions in which only two noncommutative parameters,
say $\kappa$ and $\tilde{\kappa}$, are independent,
\begin{eqnarray}
x_k(t)&=& \frac{1}{\sqrt{2}}\left[\frac{F_l}{m}\frac{{\tilde{\kappa}}^2}{F_{\gamma}^2}
-\left(x_{k0}-x_{l0}\right)+\frac{1}{2}\frac{\tilde{\kappa}}{F_{\gamma}}\left(3v_{k0}+v_{l0}\right)\right]
\sin \left(\frac{\sqrt{2}F_{\gamma}}{\tilde{\kappa}} t\right)\nonumber\\
&&-\frac{1}{4}\left[\frac{3F_k+F_l}{m}\frac{{\tilde{\kappa}}^2}{F_{\gamma}^2}
+2\left(x_{k0}-x_{l0}-2 \frac{\tilde{\kappa}}{F_{\gamma}}v_{l0}\right)\right]
\cos \left( \frac{\sqrt{2}F_{\gamma}}{\tilde{\kappa}} t\right)
\nonumber\\
&& +\frac{1}{6} \frac{F_{\gamma}^2}{\kappa\tilde{\kappa}}  t^3
-\frac{1}{4 m}\left(F_k+F_l-2 m \frac{{F_{\gamma}}}{\kappa} \right)t^2 \nonumber\\
& & + \left[\frac{F_{\gamma}}{\tilde{\kappa}}  \left(x_{k0}-x_{l0}\right)-\frac{1}{2} \left(v_{k0}+v_{l0}\right)
-\frac{F_l}{m}\frac{\tilde{\kappa}}{F_{\gamma}}\right]t\nonumber\\
&& +\frac{1}{4}\left[\frac{3F_k+F_l}{m}\frac{{\tilde{\kappa}}^2}{F_{\gamma}^2}
+2\left(3x_{k0}-x_{l0}-2 \frac{\tilde{\kappa}}{F_{\gamma}}v_{l0}\right)\right],
\end{eqnarray}
and
\begin{eqnarray}
x_l(t)&=& -\frac{1}{\sqrt{2}}\left[\frac{F_k}{m}\frac{{\tilde{\kappa}}^2}{F_{\gamma}^2}
-\left(x_{l0}-x_{k0}\right)-\frac{1}{2}\frac{\tilde{\kappa}}{F_{\gamma}}\left(3v_{l0}+v_{k0}\right)\right]
\sin \left(\frac{\sqrt{2}F_{\gamma}}{\tilde{\kappa}} t\right)\nonumber\\
&&-\frac{1}{4}\left[\frac{3F_l+F_k}{m}\frac{{\tilde{\kappa}}^2}{F_{\gamma}^2}
+2\left(x_{l0}-x_{k0}+2 \frac{\tilde{\kappa}}{F_{\gamma}}v_{k0}\right)\right]
\cos \left( \frac{\sqrt{2}F_{\gamma}}{\tilde{\kappa}} t\right)
\nonumber\\
&& +\frac{1}{6} \frac{F_{\gamma}^2}{\kappa\tilde{\kappa}}  t^3
-\frac{1}{4 m}\left(F_l+F_k+2 m \frac{{F_{\gamma}}}{\kappa} \right)t^2 \nonumber\\
& & + \left[-\frac{F_{\gamma}}{\tilde{\kappa}} \left(x_{l0}-x_{k0}\right)-\frac{1}{2} \left(v_{l0}+v_{k0}\right)
+\frac{F_k}{m}\frac{\tilde{\kappa}}{F_{\gamma}}\right]t \nonumber\\
&& +\frac{1}{4}\left[\frac{3F_l+F_k}{m}\frac{{\tilde{\kappa}}^2}{F_{\gamma}^2}
+2\left(3x_{l0}-x_{k0}-2 \frac{\tilde{\kappa}}{F_{\gamma}}v_{k0}\right)\right].
\end{eqnarray}
Note that the solutions include periodic functions and a polynomial of time to the highest power of 3.
In addition, the solution in ${\gamma}$-direction takes a simpler version that is just related to the two independent
noncommutative parameters, $\kappa$ and $\tilde{\kappa}$,
\begin{eqnarray}
x_{\gamma}(t)& = & x_{\gamma0}+\left[v_{\gamma0}+\frac{m}{\tilde{\kappa}}\left(\frac{F_{\gamma}}{\tilde{\kappa}}
\left(x_{k0}v_{l0}+x_{l0}v_{k0}\right)
-\frac{F_kx_{l0}-F_lx_{k0}}{m}
-\frac{F_{\gamma}^2}{{\tilde{\kappa}}^2}\left(x_{k0}^2-x_{l0}^2\right)\right)\right]t
\nonumber\\
& & +\left(\frac{F_{\gamma}}{m}-\frac{F_k-F_l}{\kappa}\right)\frac{t^2}{2}
-\frac{mF_{\gamma}}{{\tilde{\kappa}}^2}\Big[x_k(t)x_l(t)-x_{k0}x_{l0}\Big]\nonumber\\
& & +\int^t_0\left[\frac{F_kx_l(z)-F_lx_k(z)}{\tilde{\kappa}}-\frac{mF_{\gamma}^2}{{\tilde{\kappa}}^2}
\left[\frac{z\left(x_k(z)-x_l(z)\right)}{\kappa}
-\frac{x_k^2(z)-x_l^2(z)}{\tilde{\kappa}}\right]\right]dz.
\end{eqnarray}

As dealt with in the previous two cases, we only maintain $v_{l0}$, $F_{\gamma}$, $m$, $\kappa$,
and $\tilde{\kappa}$ nonzero, and simplify the solutions to be,
\begin{eqnarray}
x_k(t)&=&
-\frac{1}{2 \sqrt{2}}\frac{\tilde{\kappa}}{F_{\gamma}}v_{l0}\sin\left(\frac{\sqrt{2}F_{\gamma}}{\tilde{\kappa}} t\right)
-\frac{\tilde{\kappa}}{F_{\gamma}}v_{l0}
\left[1-\cos\left(\frac{\sqrt{2}F_{\gamma}}{\tilde{\kappa}} t\right)\right]
\nonumber\\
& &+\frac{1}{6}\frac{{F_{\gamma}}^2}{\kappa\tilde{\kappa}}t^3+\frac{1}{2}\frac{F_{\gamma}}{\kappa}t^2
-\frac{1}{2}v_{l0}t,\\
x_l(t)&=&
\frac{3}{2 \sqrt{2}}\frac{\tilde{\kappa}}{F_{\gamma}}v_{l0}\sin\left(\frac{\sqrt{2}F_{\gamma}}{\tilde{\kappa}} t\right)
+\frac{1}{6}\frac{{F_{\gamma}}^2}{\kappa\tilde{\kappa}}t^3-\frac{1}{2}\frac{F_{\gamma}}{\kappa}t^2
-\frac{1}{2}v_{l0}t,\\
x_{\gamma}(t) & = &
\frac{F_{\gamma}}{2m}t^2-\frac{mF_{\gamma}}{{\tilde{\kappa}}^2}x_k(t)x_l(t) \nonumber\\
& &
-\frac{mF_{\gamma}^2}{{\tilde{\kappa}}^2}\int^t_0\left[\frac{z\left(x_k(z)-x_l(z)\right)}{\kappa}
-\frac{x_k^2(z)-x_l^2(z)}{\tilde{\kappa}}\right]dz.
\end{eqnarray}
The corresponding trajectories are
illustrated in Figure 4 for typical values of the nonzero parameters, $v_{l0}$, $F_{\gamma}$, $m$, $\kappa$,
and $\tilde{\kappa}$.
\begin{figure}
\centering
\hfill
\includegraphics[width=0.32\textwidth]{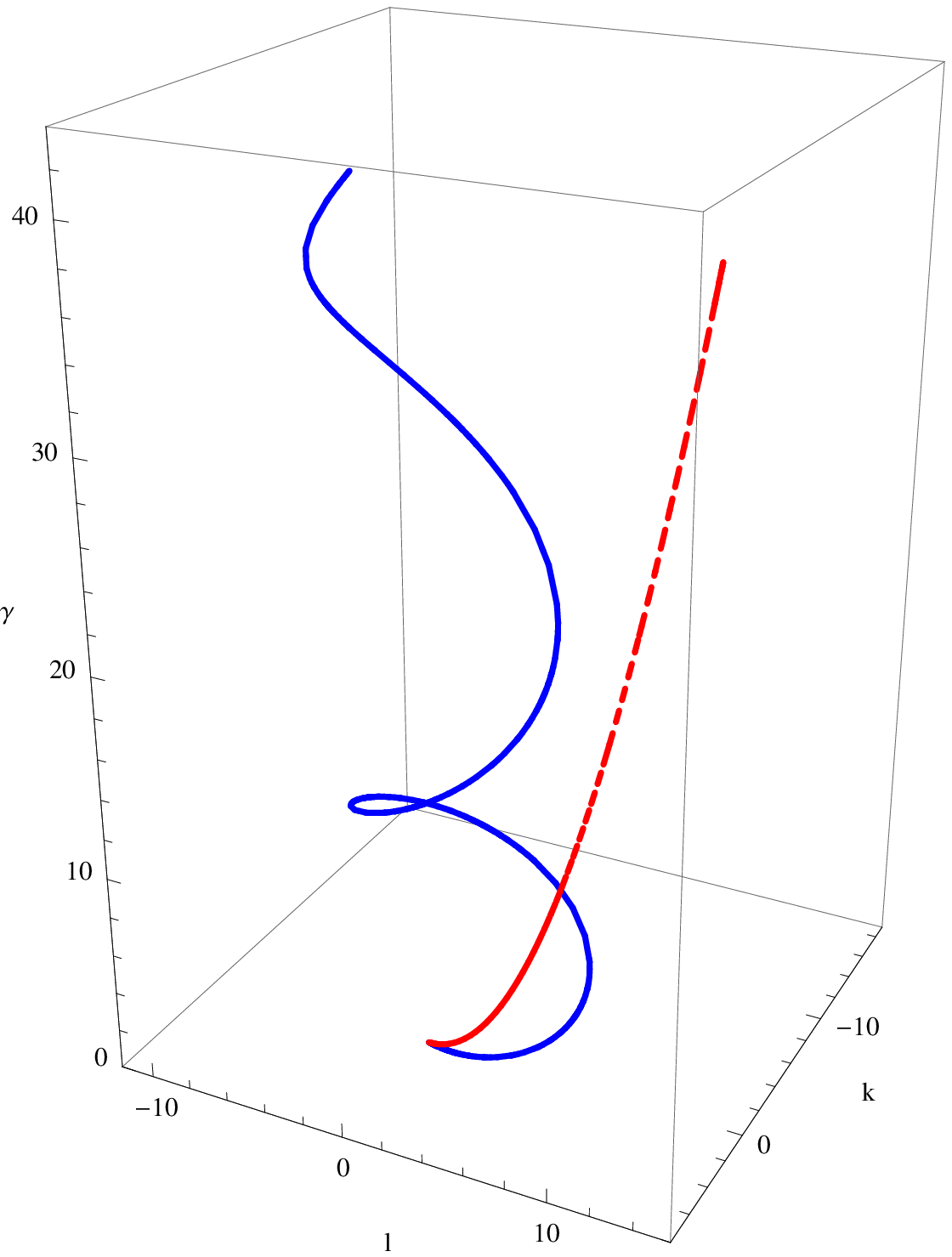}
\hfill
\includegraphics[width=0.32\textwidth]{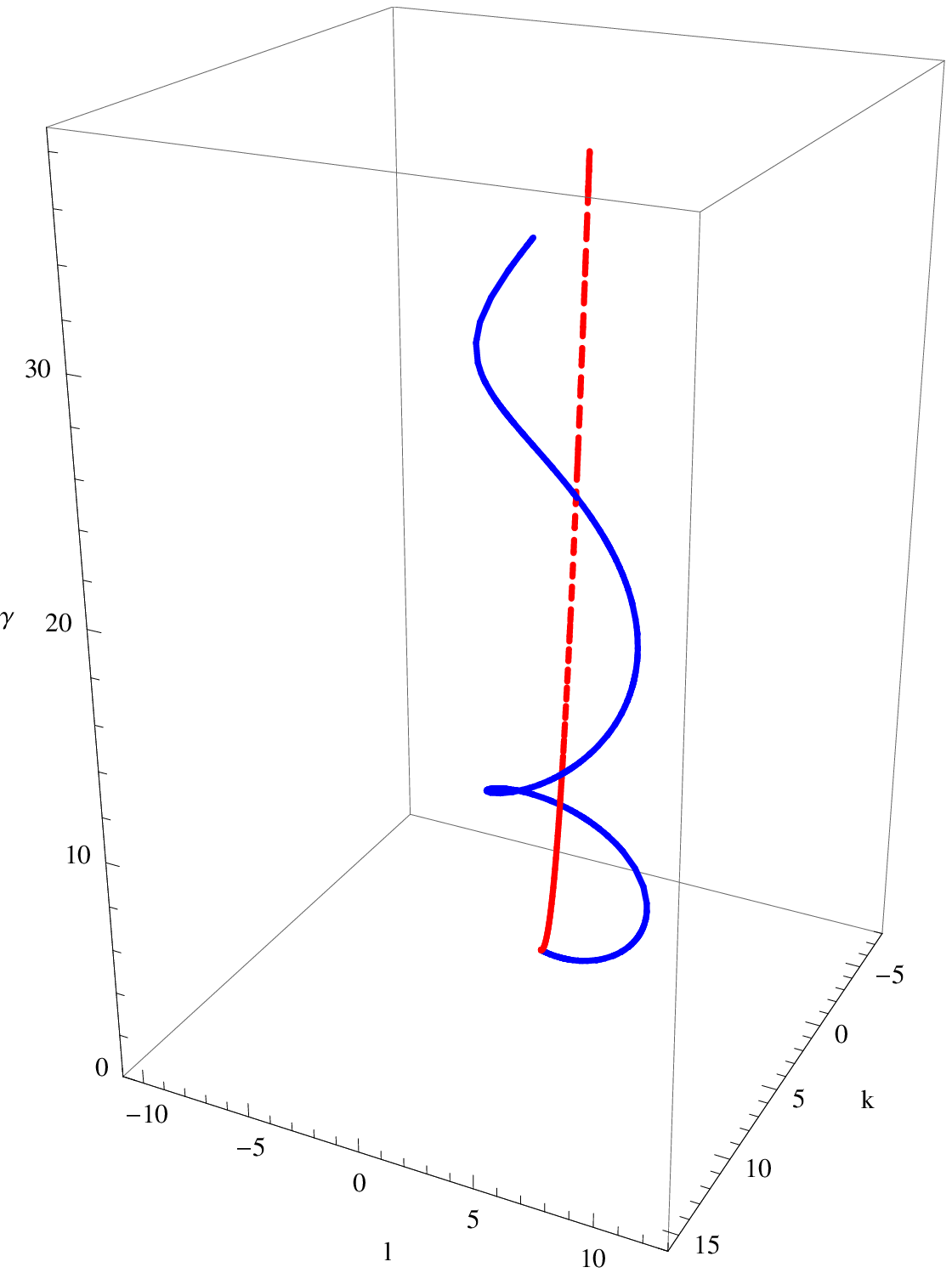}
\hspace*{\fill}
\caption{The left and right figures correspond to different noncommutative parameters
$(\kappa, \tilde{\kappa}, \bar{\kappa})=(700, 100, 10)$ and $(100, 700, 10)$,
and different $\gamma$-components of the external force
$F_\gamma=\tilde{\kappa }^2/\bar{\kappa}=(1000, 49000)$, respectively,
but to same values of the rest of parameters,
{\em i.e.} $v_{l0}=1$ and $m=1$. Note that the $\gamma$-directions of the left and right figures
are compressed by 10 and 11 times, respectively, in order to reflect the characteristic of the trajectories.
The time variable runs from 0 to $2\sqrt{2}\pi\bar{\kappa}/\tilde{\kappa}$.
To make a comparison between the deformed and undeformed
cases, the latter $(\kappa=\infty, \tilde{\kappa}=\infty, \bar{\kappa}=\infty)$ is plotted by a dash line.}
\end{figure}

\section{Conclusion}
In this paper we propose a useful method to look for new Lie-algebraic noncommutative spaces, that is,
to solve the constraint equations that the noncommutative parameters satisfy. In this way, we find out
two new Lie-algebraic noncommutative spaces which {\em not only} include that of Ref.~\cite{s5} as our special cases,
{\em but also} provide novel and strange trajectories of motion.
Based on the two types of noncommutative spaces, we study the classical mechanics of a nonrelativistic particle
interacting with a constant external force along the star-product independent way.
In accordance with the Hamiltonian analysis, we derive the equations of
motion which exhibit various marvelous extra forces arising from the noncommutativity between
different spatial coordinates and between spatial coordinates and momenta as well. In particular, we encounter
the unimaginable extra forces which are $t\dot{x}$-, $\dot{(xx)}$-, and $\ddot{(xx)}$-dependent, respectively.
Through solving the Newton equation, we obtain the various and interesting trajectories which are
extremely deformed in the way of direction-dependence by the extra forces. In addition,
we give from the point of view of trajectories of motion
the fact that the Lie-algebraic noncommutative spaces are in general anisotropic.

We note that our solutions eqs.~(6) and (7) are not the most general cases covered by the constraint
equations eqs.~(2) and (5) although they are more general than that given in Ref.~\cite{s5}. As a further consideration, we
should therefore find out more solutions of eqs.~(2) and (5) that could probably provide interesting
noncommutative spaces
with the Lie-algebraic structure. On the other hand, as a further application of our two types of noncommutative spaces
eqs.~(8) and (9), we plan to discuss some interesting systems in classical mechanics that relate to position and
velocity dependent
external forces, such as the harmonic oscillator.\footnote{The harmonic oscillator has been discussed as an example
in many noncommutative spaces different from our Type I and Type II spaces. See, for instance, some citations in
Ref.~\cite{s6}, and also see Ref.~\cite{s18} and the references therein
from the point of view of $\kappa$-deformed oscillator algebras.}
In that case, we cannot circumvent star-products. Therefore, we have to envisage the problem on how to work out the star-product
that is applicable to the algebraic structures of Type I and Type II noncommutative spaces.
Related researches are under consideration and results will be given separately.

\vspace{10mm}
\noindent
{\bf Acknowledgments}

\vspace{5mm}
\noindent
Y-GM would like to thank J. Lukierski of the University of Wroclaw and
H.J.W. M\"uller-Kirsten of the University of Kaiserslautern for helpful discussions and
H.P. Nilles of the University of Bonn for kind hospitality.
This work was supported in part by the DFG (Deutsche Forschungsgemeinschaft), by the National Natural
Science Foundation of China under grant No.10675061,
and by the Ministry of Education of China under grant No.20060055006.


\newpage
\vspace{10mm}
\baselineskip 20pt
\begin{thebibliography}{s10}
\bibitem{s1}H.S. Snyder, {\em Quantized space-time,} {\em Phys. Rev.} {\bf 71} (1947) 38;
{\em The electromagnetic field in quantized space-time}, {\em Phys. Rev.} {\bf 72} (1947) 68.

\bibitem{s2}A. Connes, M.R. Douglas and A. Schwarz, {\em Noncommutative geometry and matrix theory:
compactification on tori}, {\em J. of High Energy Phys.} {\bf 02} (1998) 003 [arXiv:hep-th/9711162];\\
N. Seiberg and E. Witten, {\em String theory and noncommutative geometry}, {\em J. of High Energy Phys.} {\bf 09}
(1999) 032 [arXiv:hep-th/9908142].

\bibitem{s3}Because of too many papers on the NCQM, we
only refer to some early works for brevity of citations.\\
G.V. Dunne, R. Jackiw and C.A. Trugenberger, {\em `Topological' (Chern-Simons) quantum mechanics},
{\em Phys. Rev.} {\bf D 41} (1990) 661;\\
J. Madore, {\em Quantum mechanics on a fuzzy sphere}, {\em Phys. Lett.} {\bf B 263} (1991) 245;\\
J. Lukierski, P.C. Stichel and W.J. Zakrzewski, {\em Galilean-invariant (2+1)-dimensional models
with a Chern-Simons-like term and D = 2 noncommutative geometry},
{\em Ann. Phys. (NY)} {\bf 260} (1997) 224 [arXiv:hep-th/9612017];\\
C. Duval and P.A. Horvathy, {\em The exotic Galilei group and the ``Peierls substitution''},
{\em Phys. Lett.} {\bf B 479} (2000) 284 [arXiv:hep-th/0002233];\\
M. Chaichian, M.M. Sheikh-Jabbari and A. Tureanu, {\em Hydrogen atom spectrum and the Lamb
shift in noncommutative QED},
{\em Phys. Rev. Lett.} {\bf 86} (2001) 2716 [arXiv:hep-th/0010175];\\
J. Gamboa, M. Loewe and J.C. Rojas, {\em Non-commutative quantum mechanics},
{\em Phys. Rev.} {\bf D 64} (2001) 067901 [arXiv:hep-th/0010220];\\
V.P. Nair and A.P. Polychronakos, {\em Quantum mechanics on the noncommutative plane and
sphere}, {\em Phys. Lett.} {\bf B 505} (2001) 267 [arXiv:hep-th/0011172];\\
P.M. Ho and H.C. Kao, {\em Noncommutative quantum mechanics from noncommutative
quantum field theory}, {\em Phys. Rev. Lett.} {\bf 88} (2002) 151602 [arXiv:hep-th/0110191];\\
R. Iengo and R. Ramachandran, {\em Landau levels in the noncommutative $AdS_2$},
{\em J. of High Energy Phys.} {\bf 02} (2002) 017 [arXiv:hep-th/0111200];\\
B. Morariu and A.P. Polychronakos, {\em Quantum mechanics on noncommutative Riemann
surfaces}, {\em Nucl. Phys.} {\bf B 634} (2002) 326 [arXiv:hep-th/0201070];\\
S. Bellucci and A. Nersessian, {\em Phases in noncommutative quantum mechanics on
(pseudo)sphere}, {\em Phys. Lett.} {\bf B 542} (2002) 295 [arXiv:hep-th/0205024].

\bibitem{s4}For brevity of citations, we just refer to several review articles that are mostly cited.\\
A. Konechny and A. Schwarz, {\em Introduction to M(atrix) theory and noncommutative geometry}, {\em Phys. Rept.}
{\bf 360} (2002) 353 [arXiv:hep-th/0012145, arXiv:hep-th/0107251];\\
M.R. Douglas and N.A. Nekrasov, {\em Noncommutative field theory}, {\em Rev. Mod. Phys.} {\bf 73} (2002) 977
[arXiv:hep-th/0106048];\\
R.J. Szabo, {\em Quantum field theory on noncommutative spaces}, {\em Phys. Rept.} {\bf 378} (2003) 207
[arXiv:hep-th/0109162].

\bibitem{s5}M. Daszkiewicz and C.J. Walczyk, {\em Newton equation for canonical, Lie-algebraic and quadratic
deformation of classical space}, {\em Phys. Rev.} {\bf D 77} (2008) 105008 [arXiv:0802.3575[math-ph]].

\bibitem{s6}J. Lukierski, H. Ruegg and W.J. Zakrzewski,
{\em Classical and quantum mechanics of free $\kappa$-relativistic systems}, {\em Ann. Phys. (N.Y)} {\bf 243} (1995) 90
[arXiv:hep-th/9312153];\\
P.A. Horvathy and M.S. Plyushchay, {\em Non-relativistic anyons, exotic Galilean symmetry and noncommutative plane},
{\em J. of High Energy Phys.} {\bf 06} (2002) 033 [arXiv:hep-th/0201228];\\
S. Benczik, L.N. Chang, D. Minic, N. Okamura, S. Rayyan and T. Takeuchi, {\em Short distance vs. long distance
physics: the classical limit of the minimal length uncertainty relation}, {\em Phys. Rev.} {\bf D 66} (2002) 026003
[arXiv:hep-th/0204049];\\
A.A. Deriglazov, {\em Poincar$\acute{e}$ covariant mechanics on noncommutative space},
{\em J. of High Energy Phys.} {\bf 03} (2003) 021 [arXiv:hep-th/0211105];\\
S. Ghosh and P. Pal, {\em $\kappa$-Minkowski spacetime through exotic ``oscillator''},
{\em Phys. Lett.} {\bf B 618} (2005) 243 [arXiv:hep-th/0502192];\\
J.M. Romero and J.A. Santiago, {\em Cosmological constant and noncommutativity: a Newtonian point of view},
{\em Mod. Phys. Lett.} {\bf A 20} (2005) 781;\\
D. Kochan, {\em Noncommutative Lagrange mechanics}, {\em SIGMA} {\bf 4} (2008) 028 [arXiv:hep-th/0610061];\\
P. Aschieri, F. Lizzi and P. Vitale, {\em Twisting all the way: from classical mechanics to quantum fields},
{\em Phys. Rev.} {\bf D 77} (2008) 025037 [arXiv:0708.3002[hep-th]];\\
P.D. Alvarez, J. Gomis, K. Kamimura and M.S. Plyushchay, {\em Anisotropic harmonic oscillator, non-commutative Landau
problem and exotic Newton-Hooke symmetry}, {\em Phys. Lett.} {\bf B 659} (2008) 906 [arXiv:0711.2644[hep-th]];\\
R. Amorim, {\em Tensor coordinates in noncommutative mechanics}, {\em J. Math. Phys.} {\bf 50} (2009) 052103
[arXiv:0804.4405[hep-th]];\\
M. Daszkiewicz and C.J. Walczyk, {\em Oscillator model on Lie-algebraically deformed nonrelativistic
space-time}, {\em Acta Phys. Polon.} {\bf B 40} (2009) 293 [arXiv:0812.1264[hep-th]];\\
M. Gomes, V.G. Kupriyanov and A.J. da Silva, {\em Dynamical noncommutativity}, arXiv:0908.2963[hep-th];\\
D.M. Gitman and V.G. Kupriyanov, {\em Gauge invariance and classical dynamics of noncommutative particle theory},
arXiv:0910.1341[math-ph].

\bibitem{s7}G. Amelino-Camelia, {\em Testable scenario for relativity with minimum length}, {\em Phys. Lett.}
{\bf B 510} (2001) 255 [arXiv:hep-th/0012238];
{\em Relativity in space-times with short distance structure governed by an observer independent
(Planckian) length scale}, {\em Int. J. Mod. Phys.} {\bf D 11} (2002) 35  [arXiv:gr-qc/0012051].

\bibitem{s8}A. Connes, {\em Noncommutative geometry}, Academic Press, New York, 1994.

\bibitem{s9}S. Zakrzewski, {\em Poisson structures on the Poincar$\acute{e}$ group}, arXiv:q-alg/9602001;\\
Y. Brihaye, E. Kowalczyk and P. Maslanka, {\em Poisson-Lie structure on Galilei group}, arXiv:math/0006167[math.QA].

\bibitem{s10}M. Daszkiewicz, {\em Canonical and Lie-algebraic twist deformations of Galilei algebra},
{\em Mod. Phys. Lett.} {\bf A 23} (2008) 505 [arXiv:0801.1206[hep-th]];
{\em Canonical and Lie-algebraic twist deformations of $\kappa$-Poincare and contractions
to $\kappa$-Galilei algebras}, {\em Int. J. Mod. Phys.} {\bf A 23} (2008) 4387 [arXiv:0802.1974[math-ph]].


\bibitem{s11}S. Giller, P. Kosinski, M. Majewski, P. Maslanka and J. Kunz,
{\em More about Q deformed Poincar$\acute{e}$ algebra}, {\em Phys. Lett.} {\bf B 286} (1992) 57.


\bibitem{s12}M. Daszkiewicz, {\em Canonical, Lie-algebraic and quadratic twist deformations of Galilei group},
{\em Mod. Phys. Lett.} {\bf A 23} (2008) 1757 [arXiv:0807.0133[hep-th]].

\bibitem{s13}R. Oeckl, {\em Untwisting noncommutative $\mathbb{R}^d$ and the equivalence of quantum field theories},
{\em Nucl. Phys.} {\bf B 581} (2000) 559 [arXiv:hep-th/0003018];\\
M. Chaichian, P. Kulish, K. Nishijima and A. Tureanu, {\em On a
Lorentz-invariant interpretation of noncommutative space-time and its implications on
noncommutative QFT}, {\em Phys. Lett.} {\bf B 604} (2004) 98 [arXiv:hep-th/0408069].

\bibitem{s14}J. Lukierski, H. Ruegg, A. Nowicki and V.N. Tolstoy, {\em q-deformation of Poincar$\acute{e}$ algebra},
{\em Phys. Lett.} {\bf B 264} (1991) 331;\\
J. Lukierski, A. Nowicki and H. Ruegg, {\em New quantum Poincar$\acute{e}$ algebra and
${\kappa}$-deformed field theory}, {\em Phys. Lett.} {\bf B 293} (1992) 344;\\
S. Majid and H. Ruegg, {\em Bicrossproduct structure of ${\kappa}$-Poincar$\acute{e}$ group and
noncommutative geometry}, {\em Phys. Lett.} {\bf B 334} (1994) 348 [arXiv:hep-th/9405107].

\bibitem{s15}J. Lukierski and M. Woronowicz, {\em New Lie-algebraic and quadratic deformations of Minkowski space from
twisted Poincar$\acute{e}$ symmetries}, {\em Phys. Lett.} {\bf B 633} (2006) 116 [arXiv:hep-th/0508083].

\bibitem{s16}J.E. Marsden and T.S. Ratiu, {\em Introduction to Mechanics and Symmetry}, Springer-Verlag, Berlin, 1999.

\bibitem{s17}G. Amelino-Camelia and M. Arzano, {\em Coproduct and star product in field theories on Lie-algebra
non-commutative space-times}, {\em Phys. Rev.} {\bf D 65} (2002) 084044 [arXiv:hep-th/0105120];\\
C. Chryssomalakos and E. Okon, {\em Star product and invariant integration for Lie type noncommutative
spacetimes}, {\em J. of High Energy Phys.} {\bf 08} (2007) 012 [arXiv:0705.3780[hep-th]];\\
S. Meljanac and S. Kresic-Juric, {\em Generalized $\kappa$-deformed spaces, star-products, and their realizations},
{\em J. Phys.} {\bf A 41} (2008) 235203 [arXiv:0804.3072[hep-th]].

\bibitem{s18}M. Daszkiewicz, J. Lukierski and M. Woronowicz, {\em $\kappa$-deformed oscillators, the choice of
star product and free $\kappa$-deformed quantum fields},
{\em J. Phys.} {\bf A 42} (2009) 355201 [arXiv:0807.1992[hep-th]].








\end{thebibliography}
\end{document}